# Characterizing the Evolution of Tropical Cyclone Thermal Structure: A Tropical Cyclone Thermal Phase Space

**Authors**: Jimin Liu[1,2,3,4], Yaoming Ma[1,3,5,6]*, Jeremy Cheuk-Hin Leung[4]*, Hong Huang[4], Shaojing Zhang[1], Zeyong Hu[2,3], Banglin Zhang[1,4,7]

**Affiliations**:

[1] College of Atmospheric Sciences, Lanzhou University, Lanzhou, China

[2] State Key Laboratory of Cryospheric Science and Frozen Soil Engineering, Northwest Institute of Eco-Environment and Resources, Chinese Academy of Sciences, Lanzhou, China

[3] University of Chinese Academy of Sciences, Beijing, China

[4] College of Meteorology and Oceanography, National University of Defense Technology, Changsha, China

[5] Land-Atmosphere Interaction and Its Climatic Effects Group, State Key Laboratory of Tibetan Plateau Earth System, Environment and Resources, Institute of Tibetan Plateau Research, Chinese Academy of Sciences, Beijing, China

[6] Kathmandu Center of Research and Education, Chinese Academy of Sciences, Beijing, China

[7] Key Laboratory of High Impact Weather (Special), China Meteorological Administration, Changsha, China

**Abstract**

The thermal structure of a tropical cyclone (TC) is a key factor in accurately forecasting its intensity. Specifically, the development, evolution, and spatial asymmetry of the warm core are closely related to TC intensification and maintenance. The current understanding of the TC warm core primarily relies on statistical averages from soundings and satellite products, which provide a climatological thermal state but cannot adequately characterize the continuous three-dimensional evolution of the warm core throughout the TC life cycle. In this study, Empirical Orthogonal Function (EOF) analysis was applied to three-dimensional temperature anomalies of Western North Pacific (WNP) TCs derived from the ERA5 reanalysis dataset for the period 1979–2024. We found that the three leading EOF modes effectively capture the primary characteristics of the TC thermal structure: the first mode represents the typical warm core structure; the second mode characterizes vertical baroclinicity; and the third pattern captures the horizontal asymmetry of the temperature anomalies. Using the first three principal components (PCs), we established a three-dimensional Cartesian coordinate system. Within this framework, two phase parameters define a thermal phase space to visualize the TC's three-dimensional thermal structure: one diagnoses the vertical barotropic or baroclinic structure, and the other quantifies the horizontal asymmetry of the thermal field. The results demonstrate that the trajectory in the phase diagram effectively captures the observed intensity changes and thermal structural evolution. The EOF-based phase diagrams offer promising insights and provide a novel, objective tool for analyzing and diagnosing structural thermodynamic characteristics and intensity evolution by quantifying key thermal features and their dynamic relationships with TC intensity. This capability thereby holds substantial potential for advancing both theoretical understanding and operational forecasting of TCs.

**Keywords** Tropical cyclone · Thermal structure · Warm core · EOF · Phase diagram

## 1 Introduction

Tropical cyclone (TC) activity has been changing significantly under greenhouse warming (Knutson et al 2019, Chand et al. 2023, Liu et al 2026). TCs are characterized by a warm core, a well-defined region of high temperature anomaly between the TC eye region and the far-field environment (Liu et al. 1999; Stern and Nolan, 2012). This structure serves as a critical diagnostic feature that distinguishes TCs from extratropical cyclones (ECs). The warm core forms as a direct thermodynamic response to the release of latent heat from the ocean through deep convection (Emanuel 1991) and subsidence warming. The warm core is directly related to TC's dynamic and thermodynamic structure through the thermal wind relationship. Thus, the evolution of the warm core is intrinsically tied to every stage of the TC's entire life cycle, from its formation and maturity to dissipation.

Given the importance of TC thermal structure, researchers have attempted to study major characteristics of TC thermal structure. The height and magnitude of the warm core are two critical parameters in studying the TC warm core. In the late 1950s and 1960s, early warm core research drew upon multilevel aircraft studies of several TCs, as noted by Stern and Nolan (2012). These limited studies demonstrated that the maximum warm core anomaly typically occurred near 250–300 hPa, often with a secondary maximum located at 600–650 hPa. Subsequent field campaigns collected a much larger soundings dataset since 1990, including aircraft-deployed dropsondes and surface-based radiosondes, which Durden (2013) used to reveal that the height of maximum warm anomaly varied between 760 and 250 hPa, and TC intensity positively correlated with both the height and magnitude of warm core. Similarly, by examining all dropsonde observations from NASA's HS3 (Hurricane and Severe Storm Sentinel) mission, Komaromi and Doyle (2017) found a very evident positive relationship between the maximum warm core temperature anomaly (regardless of height) and the intensity, a result consistent with previous studies such as Durden (2013). Meanwhile, some aircraft reconnaissance missions that penetrate the eyewall enable the study of the double warm core structure, which is closely related to TC rapid intensification (RI) (Yamada et al. 2021). With the operational application of meteorological satellites in recent decades, microwave sensors onboard these platforms, such as the Advanced Microwave Sounding Unit (AMSU) and the Atmospheric Infrared Sounder (AIRS) from NASA's Aqua satellite, have provided the capability to retrieve higher vertical resolution atmospheric temperature and humidity profiles in nearly all weather conditions since 2002. Satellite products once again confirmed the significant correlation between warm core magnitude and TC intensity in terms of minimum sea level pressure and maximum sustained wind speed (Kidder et al. 2000; Knaff et al. 2004; Zhang et al. 2007; Zhang and Chen 2012; Zhu and Weng 2013; Li and Weng 2024). Although the relationship between warm core height and TC intensity is somewhat weaker, it remains statistically significant (Gao et al. 2017). Based on 13 years of AIRS data, Wang and Jiang (2019) generated a global climatology of TC warm core and found that the typical warm core height is at the upper level around 300–400 hPa and is very similar across all ocean basins. In contrast, the warm core magnitude differs regionally: on average, the intensity is strongest (about 8 K) in the WNP and weakest (about 3 K) in the North Indian Ocean (NIO). For TCs undergoing RI, an AIRS-derived warm core magnitude greater than 4 K and a weighted warm core height higher than 450 hPa have been identified as the necessary conditions (Wang et al. 2020).

Despite the extensive statistical characteristics in TC warm core studies gained from soundings, satellite products, and reanalysis datasets, most approaches employed have largely been limited to representing the thermal field through composites, either vertical profiles of temperature anomaly or temperature anomaly on pressure levels across different TC categories within specified radii. As the TC life cycle is a continuous dynamic and thermodynamic process, wherein the evolution of its thermal structure is intrinsically linked to intensity changes and the generation of hazardous weather (Munsell et al. 2018; Tan et al. 2022; Chen and Li 2024). By their nature, however, the aforementioned composite methods only reflect an averaged thermal state within specific intensity categories. Consequently, there is a lack of effective ways to systematically characterize the continuous evolution of the thermal structure throughout the entire TC life cycle, such as during the genesis, intensification, maturity, and decay stages. This limitation restricts our in-depth understanding of the intrinsic physical mechanisms underlying how thermal structure evolution regulates TC development.

To address this issue, researchers have proposed the TC phase space as a particularly powerful framework for quantifying this structural evolution. It achieves a low-dimensional representation of the complex thermal structure through a small set of parameters with clear thermodynamic and dynamic significance from the three-dimensional atmospheric fields. These parameters, such as horizontal thermal symmetry and vertical structure features derived from geopotential height fields, enable intuitive tracking of the complete life cycle evolution trajectory of the TC thermal structure, from formation and development to dissipation, within this low-dimensional framework. Beven (1997) first proposed using phase space to analyze TC structure and defined two key parameters for this phase space: core temperature (from warm to cold) and frontal nature. Hart (2003) later advanced and expanded this concept into the cyclone phase space (CPS), which is defined by three parameters: the storm-motion-relative thickness asymmetry (which distinguishes symmetric/non-frontal from asymmetric/frontal systems), and the vertical derivative of the horizontal height gradient (which indicates cold-core versus warm-core structures via the thermal wind relationship in the upper and lower troposphere). The CPS allows for objective identification of the TC warm core solely from three-dimensional geopotential height fields in numerical weather prediction (NWP) output or reanalysis data. It can objectively classify the TC's thermal structure among tropical, subtropical, and extratropical phases, including extratropical transition (ET; Evans and Hart 2003), where a mature TC enters a baroclinic environment in mid-latitudes and transforms into an extratropical cyclone, and tropical transition (TT; Davis and Bosart 2004), where a subtropical cyclone or extratropical cyclone evolves into a TC in the presence of an upper-level disturbance. Following its introduction, the CPS was widely adopted and has been extensively applied in assessing structural predictability (Bieli et al. 2020), regional climatological statistics (Hart et al. 2006; Guishard et al. 2009; Song et al. 2011; Wood and Ritchie 2014; Zarzycki et al. 2017; Bieli et al. 2019a, b; Wang et al. 2024; Huang et al. 2024), and operational forecasting systems (Wang et al. 2023).

The CPS method enables efficient detection of the presence of warm cores and has significantly advanced the classification and analysis of tropical cyclone structure evolution. However, the CPS approach offers only a partial view of the TC thermal structure. For example, the storm-motion-relative thickness asymmetry parameter, which is confined to the 600–900 hPa levels, more directly reflects asymmetry in the lower troposphere and thus cannot capture the thermal structural symmetry in the upper troposphere. In order to address these limitations and systematically reveal the fundamental characteristics of TC thermal structure, we aim to construct another phase space that solely takes the TC thermal structure into account, which we refer to as the TC thermal phase space. To achieve this, it is first necessary to objectively and quantitatively extract dominant modes of variation from massive meteorological datasets. Empirical Orthogonal Function (EOF) analysis serves as an ideal tool for this purpose, as it can extract dominant orthogonal modes directly from the three-dimensional temperature fields and still offer an objective, low-dimensional representation of the thermal structure (Wilks 2019; Du et al. 2025). The primary objective of this study is to demonstrate the potential of using the EOF method for capturing TC's thermal structure, an approach that has not yet been validated in this aspect of studies. In addition, we aim to employ the EOF method to construct a novel thermal phase space that enables the objective monitoring and tracking of the evolution of TC thermal structures' life cycles according to distinct evolutionary trajectories. The remainder of this study is organized as follows. Section 2 describes the data and methodology. Section 3 identifies the dominant modes of temperature anomalies in the WNP TCs and uses two TC case studies to illustrate how the PCs time series relate to TC intensity changes. Section 4 defines a thermal phase space using two parameters derived from the PCs. By analyzing their distribution patterns from 1979 to 2014, we systematically identify the representative thermal structures across different sections of the phase space. Finally, Section 5 summarizes our findings.2 Data and methods

### 2.1 Best track and reanalysis data

The observed TC data were taken from the International Best Tracks Archives for Climate Stewardship (IBTrACS, V04r01, Knapp et al. 2010, 2018). Storm tracks, wind intensities every 6 hours during the storm's lifetime for the period 1979–2024 were employed in this study. To ensure the accuracy of the analysis, we discarded all interpolated longitude, latitude, and intensity estimates and retained the original observational records. Only TCs reaching at least tropical storm strength (maximum sustained winds $\geqslant$34 knots) over the WNP are selected. The air temperature is extracted from the fifth generation of the European Centre for Medium-Range Weather Forecasts global atmospheric reanalysis dataset (ERA5) (Hersbach et al. 2020) on a 0.25°×0.25° spatial grid and 37 standard pressure levels. Reanalysis datasets have been widely used to study the TC warm core structure (Yang et al. 2016; Gao et al. 2018; Xi et al. 2021; Niu et al. 2021), investigating key aspects such as its vertical structure and robust correlation with TC intensity.

From 1979 to 2024, a total of 1,186 TC cases with 24,472 records were utilized, including 451 tropical storms (TSs), 198 Category 1 (CAT1) TCs, 130 Category 2 (CAT2) TCs, 94 Category 3 (CAT3) TCs, 185 Category 4 (CAT4) TCs, and 128 Category 5 (CAT5) TCs.

### 2.2 EOF analysis of TC temperature anomaly

The empirical orthogonal function (EOF) method has been extensively used in atmospheric and oceanic analyses as a useful technique to decompose a signal varying spatiotemporally into separated statistical modes. (Zhao et al. 2020; Song et al. 2021; Leung et al. 2022; He et al. 2025). Generally, the temperature anomaly was defined as the difference between the temperature and a reference environmental temperature at a certain pressure level. Previous studies have shown that a universally accepted definition representing the reference environment is not yet available (Stern and Nolan 2012; Munsell et al. 2018). Among the various definitions proposed, a commonly utilized method for representing environmental temperature involves calculating the mean temperature over an annulus within a specified range of distances from the storm center (Halverson et al. 2006; Zhu and Weng 2013). For this analysis, the temperature anomaly is obtained in two steps: first, for each synoptic time, we remove the 30-year climatological mean (1991–2020, recommended by WMO, 2023) for the corresponding calendar day from the original ERA5 temperature field, and then subtract the azimuthal average within an annulus 600–1000 km from the TC center.

The temperature anomaly field with a resolution of 0.25°×0.25° is interpolated to a 25km×25km equal spacing grid using the spline interpolation method. In the EOF analysis, we select a 1000 km×1000 km area centered on the TC, because a 500 km radius is generally sufficient to encompass the full-scale structure of the system, specifically the inner core and outer bands (Zhu et al. 2024). On the other hand, this choice avoids treating unrelated large-scale environmental fields, such as the subtropical high and mid-latitude troughs, as part of the TC itself due to an excessively large radius (Stern and Zhang 2016). The vertical range is selected from 150 to 850 hPa, with an interval of 50 hPa, covering a total of 15 vertical levels. Atmospheric levels below 850 hPa are susceptible to influences from land-sea contrasts, topography, and surface turbulence, whereas those above 150 hPa may be affected by stratospheric cold air intrusions (Rivoire et al. 2016). These external factors are not directly associated with the TC itself, which is primarily maintained by non-adiabatic heating in the eyewall and subsidence warming in the eye. Consequently, these levels should be excluded from the analysis. Typically, the maximum temperature anomaly in a TC, also known as the warm core, is observed at 200–400 hPa. Considering that this study focuses on the thermal structure of TC and follows the common practice in TC dynamics diagnostics, we define the 150–400 hPa levels as the mid-to-upper levels, the 400–600 hPa levels as the mid levels, and the 600–850 hPa levels as the lower-to-mid levels.

For the three-dimensional temperature anomaly field at each time, the PCs obtained from the EOF decomposition represent the temporal weights or amplitudes of the original temperature anomaly field in the corresponding EOF modes. The larger the absolute value of the PCs, the more dominant the corresponding EOF mode is in the temperature anomaly field at that time, and the stronger the signal. Typically, the first $N$ modes with cumulative variance contribution rates exceeding 90% are selected. These $N$ modes retain most of the information of the original temperature anomaly field, eliminating minor noise or small-scale disturbances and laying the foundation for analyzing the "main signal." Here, we calculate the proportion of the square of the $i$-th PC to the total sum of the squares of all $N$ PCs, that is $PC_i^2 / \sum_{i=1}^{N} PC_i^2$, which can quantitatively measure the relative importance of the $i$-th mode at a particular time.

### 2.3 Definition of TC stages

The TC intensity change is typically quantified by the 24-hour change in maximum sustained winds, which is defined as $\Delta V_{24} = V(t+24) - V(t)$, where $V$ is the maximum sustained wind speed. Following Jiang and Ramirez (2013), we classify TC intensity changes into four categories based on the value of $\Delta V_{24}$: Weakening (W, $\Delta V_{24}$<10 knots), Neutral (N, -10 knots ≤ $\Delta V_{24}$ < 10 knots), Slowly Intensifying (SI, 10 knots ≤ $\Delta V_{24}$ < 30 knots) and Rapid Intensifying (RI, $\Delta V_{24}$ ⩾ 30 knots).

## 3 Dominant modes of TC thermal structure

### 3.1 The three leading EOF modes of TC temperature anomalies

In order to unravel how the TC's temperature anomaly field evolves, it is necessary to capture its most representative spatial pattern. Thus, in this section, we first applied the EOF analysis to the temperature anomaly field to extract the dominant modes and corresponding PCs. As shown in Fig.

1, temperature anomaly in TC over the WNP basin during 1979–2024 can be represented by three distinct EOF patterns, explaining 61.1% of the total variance. Specifically, the three leading EOF modes explains 47.8%, 8.3% and 5.0% of the variances, respectively. Three dominant EOF modes are significantly statistical separation by eigenvalue following the rule of thumb proposed by North et al. (1982).

The first leading EOF (EOF1) mode (Fig. 1a) represents the most typical characteristic of the TC warm core. The positive temperature anomalies are present throughout the 150–850 hPa levels, and the warm core itself is situated between 250–400 hPa. A notable feature is that the temperature anomaly is considerably more expansive horizontally in the mid levels and mid-to-upper levels compared to the lower-to-mid levels. Thus, we name the EOF1 mode as the "typical warm core mode".

Fig. 1b displays the second EOF (EOF2) mode, which is identified by its vertical asymmetry, demonstrating positive anomalies above the 500 hPa in contrast to negative anomalies below, which reveals a pronounced baroclinic structure. The center of positive anomalies is observed at 250–300 hPa, exhibiting a weaker intensity yet a higher altitude compared to the warm core in Fig. 1a. Furthermore, the centers of negative anomalies in the lower-to-mid levels are not exactly collocated with the TC center but are instead distributed on either side of it. We name the EOF2 mode as the "baroclinic mode", which captures the $1^{st}$ order baroclinic feature of the temperature anomaly.

The third (EOF3) mode, presented in Fig. 1c, exhibits a distinct horizontal asymmetrical pattern. Below 200 hPa, negative anomalies dominate the western flank of the TC with a center around 600 hPa, while the eastern positive anomalies are slightly weaker. Above 200 hPa, the environment is dominated by positive anomalies, with negative anomalies confined to a small area. We name the EOF3 mode as the "horizontal asymmetry mode".

Since the EOF decomposition was performed exclusively on temperature anomaly fields at or above TS intensity, no direct PCs are available for tropical depressions (TDs) or weaker stages. Thus, the corresponding PCs for these non-TS-above stages were derived via projection onto the precomputed EOF modes. In this study, we retain the first 100 EOF modes and analyze the corresponding PCs of the temperature anomaly field at each time step. The cumulative variance contribution rate of the first 100 modes is 91.9%, which meets the threshold requirement of 90%. The collective contribution of the first three EOF modes at each time was computed, that is $R_{123} = \sum_{i=1}^{3} PC_i^2 / \sum_{i=1}^{N} PC_i^2$ . The magnitude of this value serves as a direct proxy for how well the leading modes capture the essential features of the original temperature anomaly field.

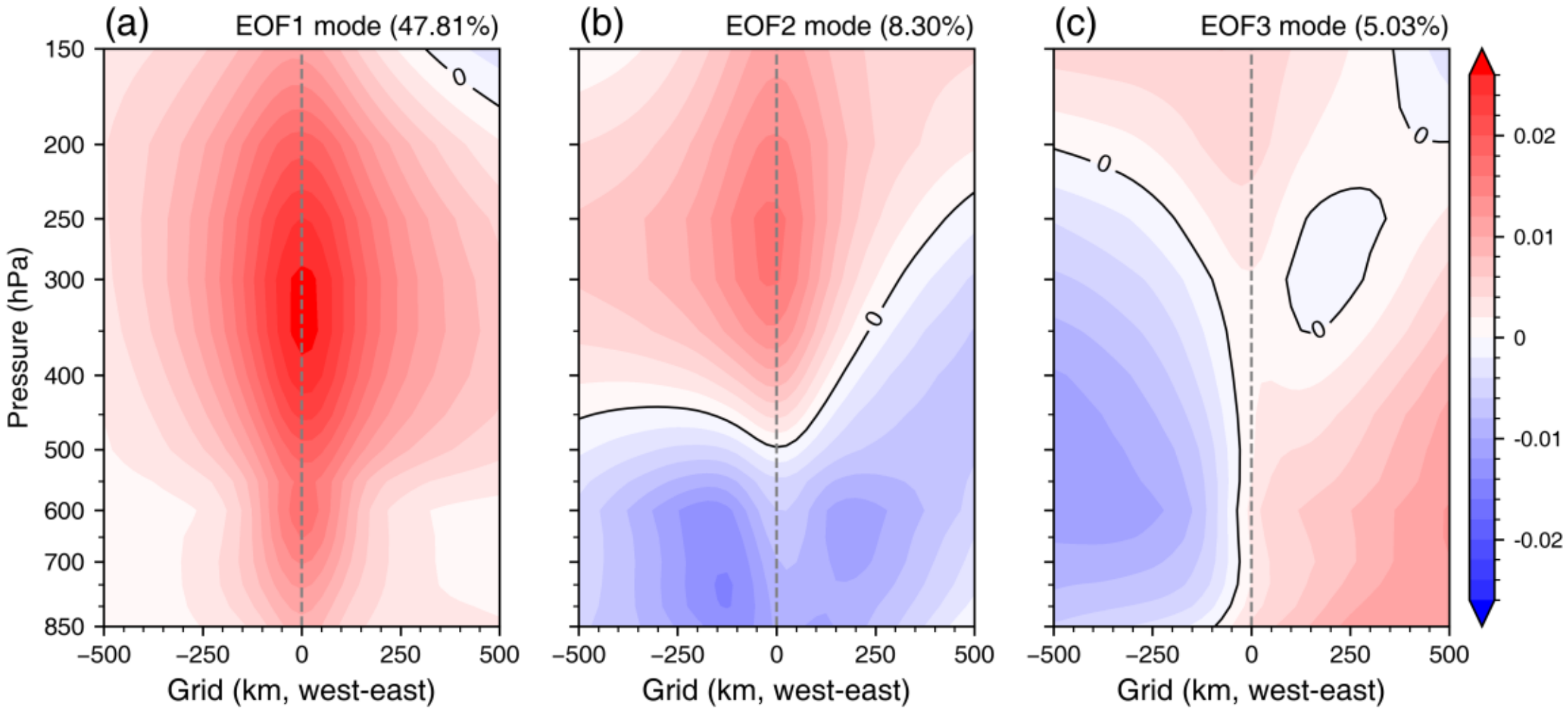

**Fig. 1** The first three dominant EOF modes of TC temperature anomaly over WNP during 1979–2024 with their corresponding explained variance (%) marked in parentheses. The black solid line denotes the zero contour. The vertical dashed line denotes the location of the TC center.

Fig. 2 shows boxplot of $R_{123}$ values for different TC categories. As TC intensity increases, the sample frequency decreases markedly from 12,858 cases at TS intensity to only 535 at CAT5. The average $R_{123}$ values for the TS, CAT1, CAT2, CAT3, CAT4, and CAT5 categories were 49.8%, 61.2%, 64.2%, 66.4%, 68.5%, and 70.4%, respectively. A monotonic increase in the mean $R_{123}$ is observed with rising intensity, accompanied by a reduction in $R_{123}$ variance. These results indicate that stronger TCs are more dominantly explained by the third leading EOF modes introduced above. This is mainly because stronger TCs are associated with more intense positive temperature anomalies, a more pronounced warm core near the storm center, and thus greater dominance of PC1. Under such conditions, the first three EOF modes can explain the TC temperature anomaly field as effectively as possible at this time. However, when the $R_{123}$ is sufficiently low, the first three EOF modes no longer represent the primary structure of the warm core anomaly. This typically occurs when a tropical cyclone is relatively weak, as the warm core structure is often disorganized and highly thermally asymmetric during its formative or decaying stages (Emanuel 1991).

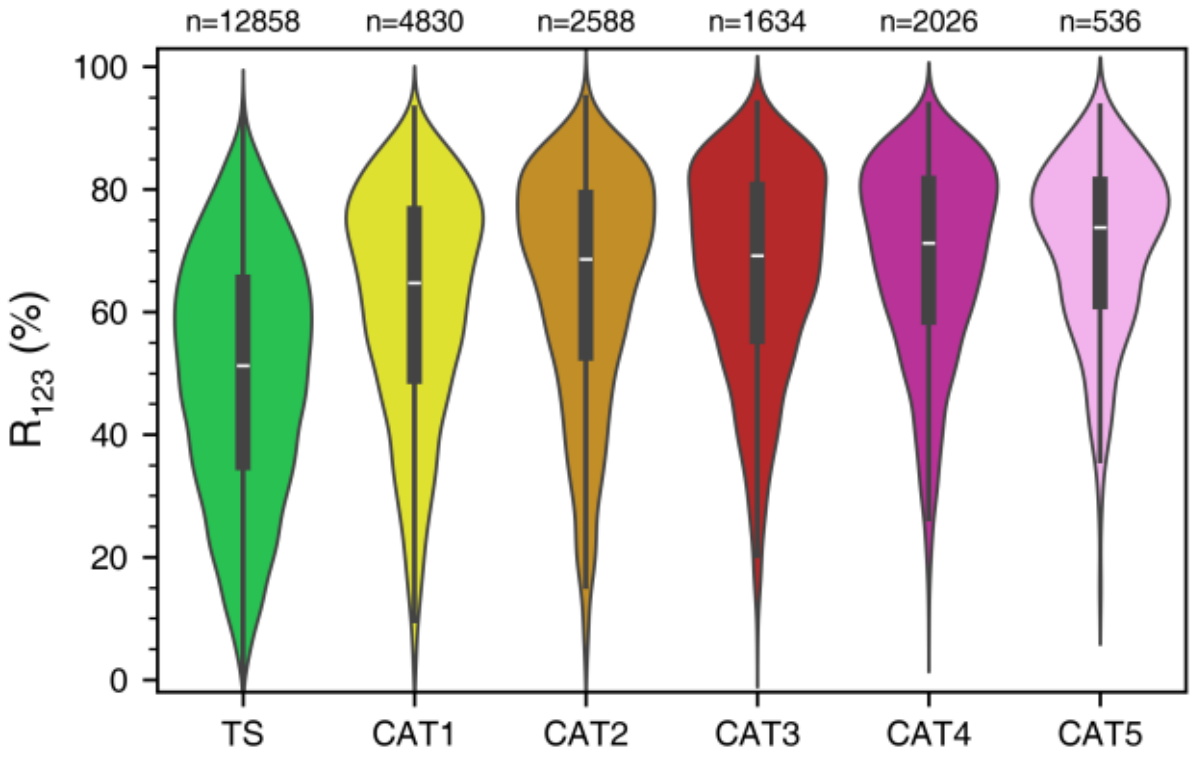

**Fig.2** Violin plot of $R_{123}$ (units: %) for different categories. The violin plots show the whole statistical distribution. The black boxes extend from the first to the third quartile, and the white line represents the median value. The sample size for each category ($n$) is shown above the figure, respectively.

### 3.2 Representation of TC thermal structure evolution based on PCs

Based on the above analyses, three dominant modes of TC temperature anomaly are identified over the WNP: a barotropic structure with uniform-sign anomaly (the typical warm core mode, EOF1), a baroclinic structure with opposing anomalies in the vertical (the vertical asymmetry mode, EOF2), and an asymmetric structure with a negative (positive) in the west (east) (the horizontal asymmetry mode, EOF3). The EOF modes can represent the evolution of the temperature anomaly field throughout the TC life cycle and are intrinsically linked to atmospheric and oceanic dynamic and thermodynamic processes. To illustrate this, we will analyze two specific TC cases, Chaba (2022) and Kong-rey (2024), to demonstrate how the PCs time series of each mode indicate the evolution of a TC's thermodynamic structure.

#### 3.2.1 Chaba (2022)

Chaba was the third TC of the 2022 WNP season and the first to make landfall over mainland China. Chaba formed over the South China Sea at 1200 UTC on 29 June 2022 and was upgraded to a TS at 06:00 UTC on 30 June (Fig. 3a). Moving northwestward subsequently, it intensified further, reaching its peak intensity of 75 knots (Category 1) at 0000 UTC on 2 July (Fig. 3b). The system then made landfall on the coast of Guangdong Province at 0600 UTC the same day, after which it moved deeper inland and weakened, and it was downgraded to a TD at 1200 UTC on 3 July. It brought strong winds and heavy downpours to the regions along the way. Chaba exhibited a classic characteristic of TCs originating in the South China Sea, namely an asymmetric structure which led to an asymmetric distribution of wind and rain (Xu et al. 2014).

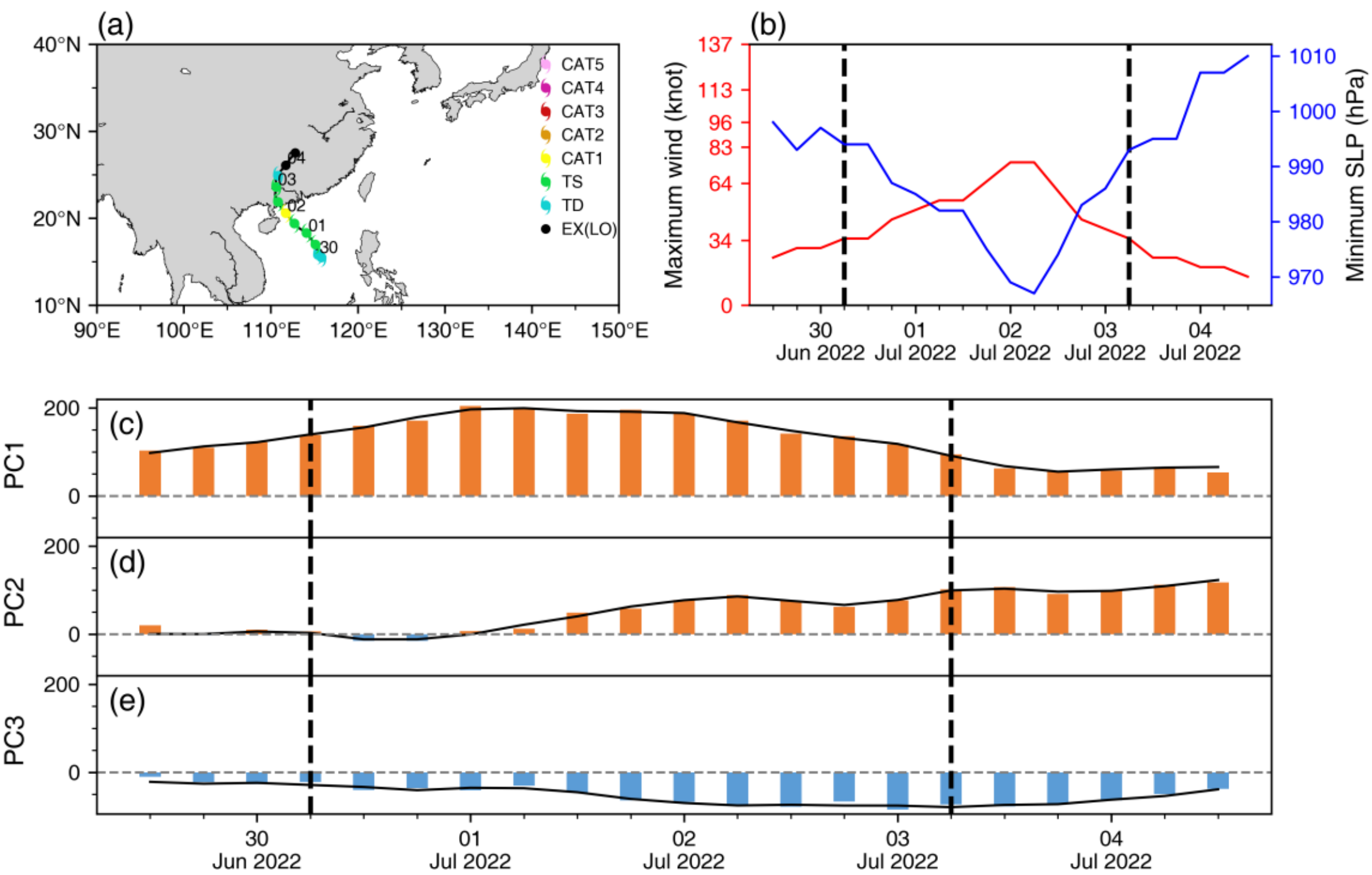

**Fig. 3** (**a**) Track and intensity of Chaba (2022). (**b**) Time series of maximum sustained wind speed (units: knots, red) and minimum sea level pressure (units: hPa, blue). (**c**)–(**e**) Bar plots of the time series of PC for the first three modes, with the black solid line representing the 5-point smoothed curve. In (a), the track colors indicate TC intensity: TD (Tropical Depression), TS (Tropical Storm), Cat1–Cat5 (Hurricane Categories 1–5), EX (Extratropical System), and LO (Low). In (**b–e**), the two vertical black dashed lines denote the first and last times at which Chaba reached tropical storm intensity.

During the initial development of Chaba (prior to 1200 UTC on 1 July), PC1 plays a dominant role (Fig. 3c), PC2 is close to 0 (Fig. 3d), and PC3 has a small negative value (Fig. 3e), indicating that the positive temperature anomaly center of the TC is in the mid-to-upper levels, and the temperature on the western side of the lower-to-mid levels is higher than that on the eastern side. Here, we examine the temperature anomaly field when Chaba first reached TS intensity at 0600 UTC on 30 June 2022 to illustrate Chaba's thermal structure as the initial stage. At this time, the warm core was weak and its anomaly maximum was situated at 400–500 hPa (Fig. 4a, c) and exhibited a westward tilt with height. At 250 hPa (Fig. 4b), the anomaly center had shifted approximately 200 km southwest of the TC center. At 750 hPa (Fig. 4d), positive values are primarily concentrated on the western side of the TC center, while negative values dominate the eastern side, this pattern at 750hPa is consistent with PC3 variations in Fig. 3e.

As the TC developed, intense updrafts within the eyewall transported abundant warm, moist air from the lower layers upward. The subsequent release of latent heat through condensation in the mid-to-upper levels warmed the core, then the vertical height of the warm center raised. When Chaba reached its lifetime maximum intensity (LMI) at 0000 UTC on 2 June, the warm core center was located at 250 hPa (Fig. 4e) and had shifted closer to the center (Fig. 4f). Moreover, the positive temperature anomalies became more vertically aligned, indicating a stronger barotropic structure. At 750 hPa (Fig. 4h), the temperature anomaly field appeared disorganized without a distinct warm core. The area of negative anomalies on the eastern side of Chaba increases, with the inner core (<200 km of the TC center; Zhu et al. 2024) nearly surrounded by negative values. Compared with the initial stage, the warm core in the mid-to-upper levels has intensified, while negative anomalies and increased asymmetry have emerged in the lower-to-mid levels. This evolution is accurately captured by the PCs: a strengthening of PC1, coupled with a notable increase in PC2 and a steady negative value in PC3 (Fig. 3c–e).

At 0600 UTC on 3 June 2022, the last time Chaba maintained TS intensity after reaching its LMI (Fig. 4i), the warm core in the mid-to-upper levels was still located around 250 hPa in the vertical structure, but its intensity weakened and the area of large positive anomalies shifted to the west (Fig. 4j). The vertical extent of the negative anomaly region further expanded, with the positive anomaly area at 500 hPa concentrated in the northwest quadrant of the TC center (Fig. 4k), and the remaining areas were mostly negative, indicating a more pronounced asymmetry (Fig. 3e). At the 750 hPa of lower-to-mid levels, there were basically all negative anomalies, which was related to the continuous increase of PC2 (Fig. 4l). Subsequently, during the TD stage, the warm core further weakened as the lower-to-mid levels were invaded by negative anomalies. The system ultimately lost its tropical characteristics and transitioned into an extratropical cyclone (Fig. 3a, d). Evidently, the decrease in PC1 during the decaying phase well captures the weakening of the mid-to-upper levels warm core, while the concurrent increase in PC2 indicates the gradual dominance of negative anomalies in the lower-to-mid levels. The negative values of PC3, corresponding to the westward shift of positive anomalies, are consistent with the warm center being located to the west of the TC center (Fig. 4i, j, k).

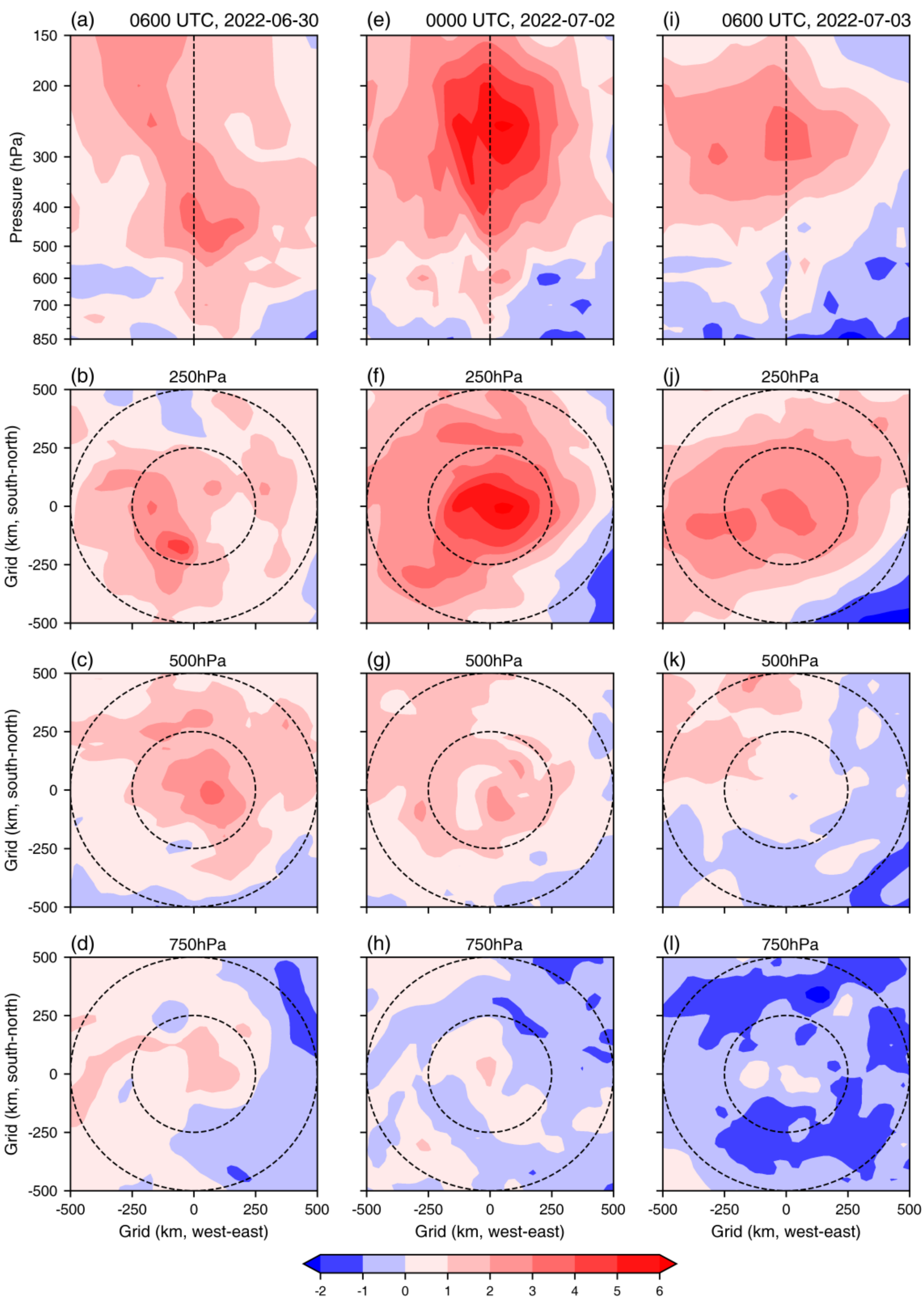

**Fig. 4** Temperature anomaly (units: °C) derived from the ERA5 dataset for Chaba, (**a–d**) zonal cross section along the TC center and horizontal panels at 250, 500, and 750hPa for 0600 UTC on 30 June 2022. (**e–h**) as in (**a–d**) but for 0000 UTC on 2 July 2022, (**i–l**) as in (**a–d**) but for 0600 UTC on 3 July 2022. The black dashed line in the first row indicates the TC center from IBTrACS. The black dashed circles in horizontal panels represent radii of 250km and 500km from the TC center, respectively.

### 3.2.2 Kong-rey (2024)

Kong-rey formed near Guam at 0000 UTC on 25 October 2024 (Fig. 5a). Following a period of slow development, it underwent RI under favorable environmental conditions, with the maximum sustained wind speed reaching 140 knots (Category 5) at 0000 UTC on 30 October (Fig. 5b). It subsequently made landfall on Taiwan Island at 0600 UTC on 31 October, and then weakened. It subsequently turned northeastward into the East China Sea and finally transitioned into an extratropical cyclone at 1200 UTC on 1 November.

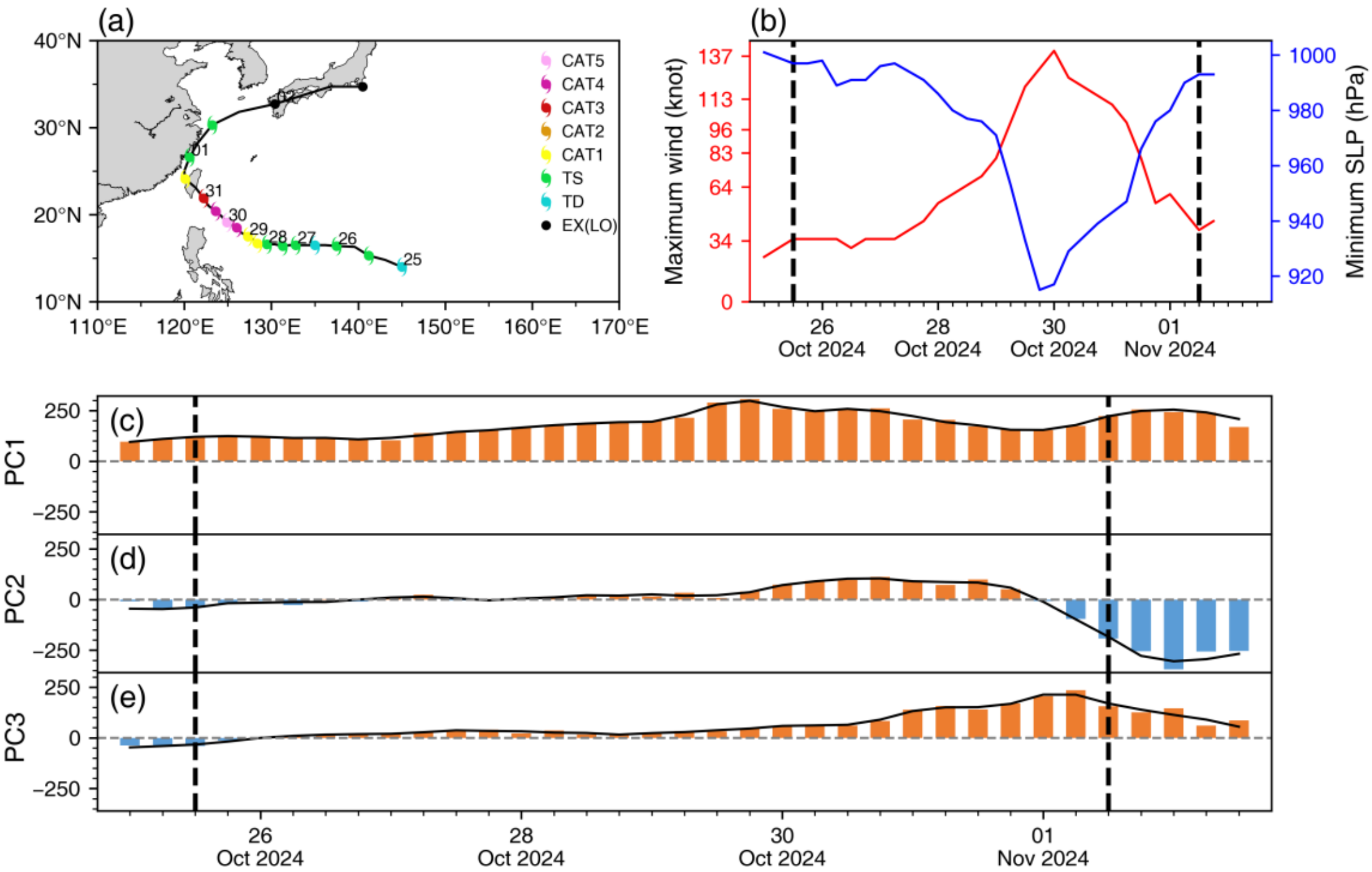

**Fig. 5** The same as **Fig. 3**, but for Kong-rey (2024).

Unlike Chaba, which formed in the South China Sea, Kong-rey developed over the vast ocean near Guam with sufficiently warm sea surface temperatures and low wind shear. This difference in genesis location led to distinct steering influences: Chaba, being farther from the western Pacific subtropical high (WPSH), experienced weaker steering flow and was more susceptible to the southwest monsoon (Wang and Zhou 2024). Conversely, Kong-rey was more influenced by the WPSH, moving northwestward under its steering flow. After recurving near the western ridge point of the WPSH, it had a greater potential to interact with mid-latitude systems such as upper-level trough and the westerly jet stream (Evans et al. 2017), resulting in a thermal structure fundamentally different from that of Chaba.

Before Kong-rey reached LMI, PC1 dominated absolutely, while the values of PC2 and PC3 were close to 0 (Fig. 5c–e). This was because Kong-rey formed over the warm ocean at low latitudes under an approximately barotropic environment. In the initial stage, at 0000 UTC on October 27, the temperature anomaly field was relatively weak, with the maximum value centered in the 300–400 hPa layer (Fig. 6a). At 250 hPa (Fig. 6b), there was scattered convection in the inner core, with unorganized positive temperature anomaly large values. The positive maximum value appeared on the west side of the TC center, about 250 km away. In contrast, at 500 hPa (Fig. 6c), the positive anomaly area was situated in the northeast quadrant of the TC center. This vertical misalignment of anomaly centers suggests that Kong-rey had not yet developed a symmetrically organized cloud structure, with convective activity still being relatively fragmented. At 750 hPa (Fig. 6d), negative anomalies prevailed west of the center beyond 250 km, while positive anomalies were found to the east. This may be because, in the weaker stage of Kong-rey, the east side of the TC was close to the WPSH. The adiabatic warming within the subtropical high subsidence raised the temperature of the boundary-layer inflow entering the TC. Between 0600 UTC on 28 October and 0600 UTC on 29 October, the 24-hour change in maximum sustained wind speed of Kong-rey met the threshold for RI, and the RI process lasted for more than 24 hours. During the RI process, convective bursts developed in the eyewall (Yang et al. 2019), and latent heat release in the mid-to-upper levels made the warm core stronger. These results indicate that when Kong-rey intensifies over the vast ocean in the lower latitudes, whether undergoing gradual development or RI, it is less affected by other weather systems and exhibits a more barotropic structure. Under these conditions, PC1 dominates (Fig. 5c), with positive temperature anomalies occupying most of the domain and horizontal asymmetry remaining weak, as evidenced by the negligible magnitudes of PC2 and PC3.

Following the RI, Kong-rey reached its LMI at 0000 UTC on 30 October, with maximum wind of 140 knots and a minimum sea level pressure of 927 hPa (Fig. 6b). The thermal structure at LMI was vertically aligned, and a double warm core structure was observed: an upper warm core centered in the 250–350 hPa layer with a maximum anomaly exceeding 10°C, and a lower warm core near 600–700 hPa with a maximum anomaly of about 8°C, which was weaker than the upper core (Fig. 6e). Consistent with this structural evolution, the value of PC1 reached its maximum after RI, while both PC2 and PC3 remained small in comparison (Fig. 5c, e). A marked increase in PC2 and PC3 was observed following Kong-Rey's LMI (Fig. 5d, e), a response directly linked to the TC's approach and eventual landfall on Taiwan (Fig. 5a). This landfall-induced abrupt cut-off of surface fluxes disproportionately distorts the lower-to-mid levels warm core, thereby amplifying the thermal asymmetry (Tsujino and Tsuboki 2020). The PC series well captures the evolution of Kong-rey's thermal characteristics during its landfall.

As Kong-rey curved northwestward, it encountered a mid-latitude upper-level trough in its northwestern (not shown), allowing cooler air to gradually intrude into the mid-to-upper levels of the TC. At 1800 UTC on 1 November, the last timestep at which Kong-rey maintained TS intensity, the warm core had collapsed, and positive anomalies were mainly concentrated below the 250 hPa (Fig. 6i). The warm core above the TC center was no longer sustained, with the positive anomaly area expanding horizontally and its maximum moving away from the center. At 250 hPa (Fig. 6j), the western side of the TC was dominated by negative anomalies, while positive anomalies persisted to the east, indicating the loss of a strong, centralized warm core. Similarly, at 500 hPa (750 hPa), the warm core of the TC center disappeared, with the positive anomaly area concentrated in the northeast quadrant (north side) of the TC (Fig. 6k, i), respectively. Similarly, at 500 hPa and 750 hPa, the warm core had vanished, with positive anomalies concentrated in the northeastern and northern quadrants, respectively (Fig. 6k, l). During the decay and extratropical transition, PC2 underwent the most pronounced change, shifting rapidly from positive to negative values. In the TD stage, the absolute value of PC2 exceeded that of PC1 (Fig. 5c, d), indicating that the mid-to-upper levels warm-core structure had been disrupted, while a residual warm core anomaly persisted in the lower-to-mid levels, maintaining horizontal asymmetry in the temperature anomaly field.

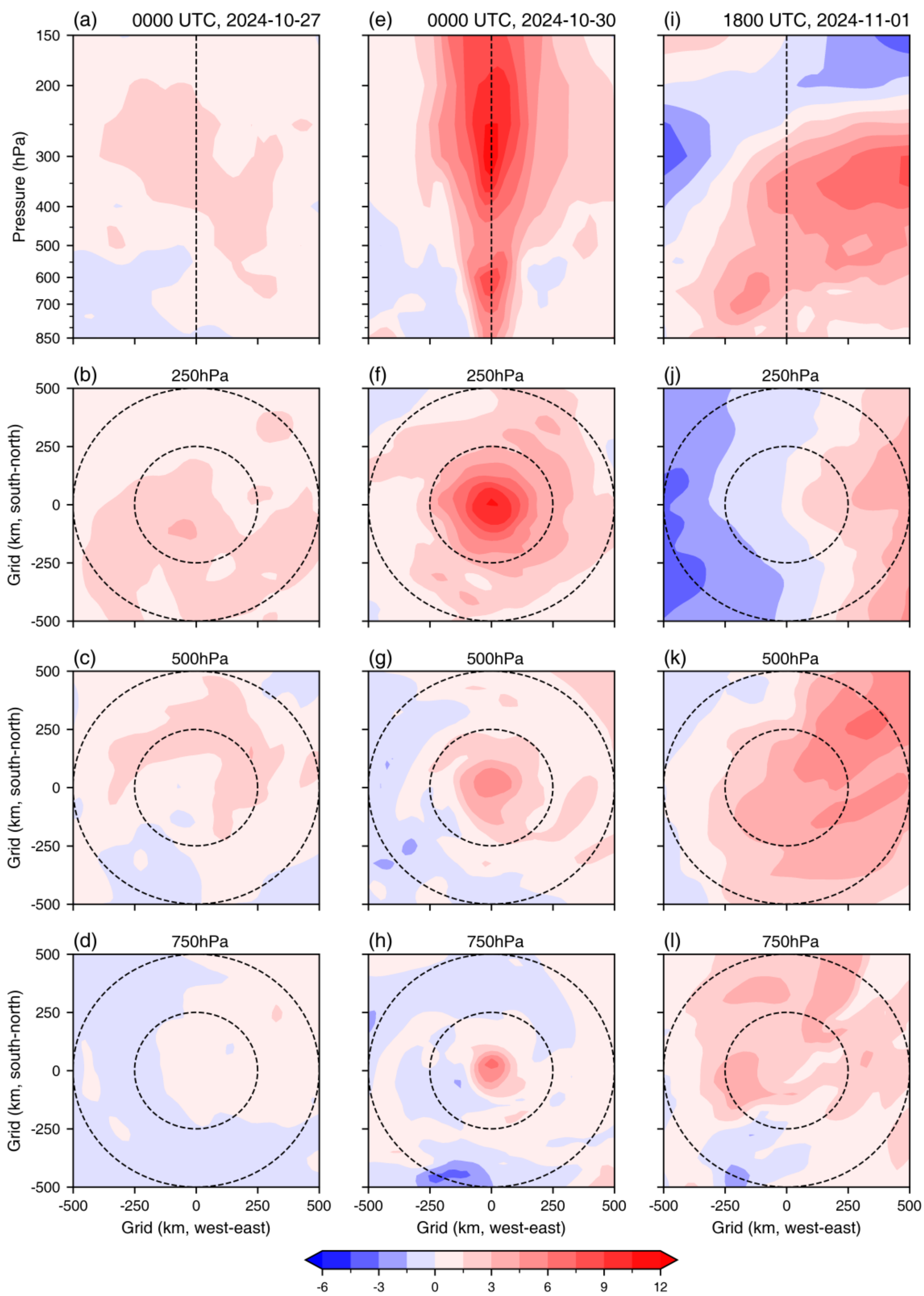

**Fig. 6** The same as **Fig. 4**, but for Kong-rey (2024).

## 4 TC thermal phase space

In Section 3, we have illustrated that the first three PCs corresponding to the EOF modes can effectively represent the major thermal characteristics of TCs, including the thermal anomalies and structural changes during the TC lifetime and extratropical transition. This suggests that one can quickly get a sense of a TC's thermal structure simply based on the PCs of the three leading EOF modes. Therefore, we attempted to use the statistical features of these three PCs to construct a thermal phase space that can visualize a TC's primary thermal structure and its evolution. The trajectory of TC in this phase space diagram can capture changes in the intensity of the warm core, vertical thermal anomalies, and horizontal thermal asymmetry throughout the TC's life cycle. In the following, we first introduce how we construct this thermal phase space, and then demonstrate how this thermal phase space can be interpreted and applied to TC analyses.

### 4.1 Definitions

A Cartesian coordinate system is established with the three PCs as axes (Fig. 7a). The position of any point *P* in this coordinate system can be expressed as *P* (PC1, PC2, PC3). The plane formed by PC1 and PC2 is denoted as the PC1–PC2 plane, and that formed by PC1 and PC3 is denoted as the PC1–PC3 plane, and so on. The two parameters describing the thermal structure of the TC are defined as the angles formed by the PCs: the angle ϕ between point P and the PC1–PC2 plane, and the angle λ between the projection of point P onto the PC1–PC2 plane and the PC1 axis. Both ϕ and λ are expressed in degrees, with ϕ ranging from -90° to 90° and λ ranging from -180° to 180°. The corresponding formulas are given as follows:

$$\begin{aligned} \tan\lambda &= PC2 / PC1 \\ \tan\phi &= PC3 / \sqrt{PC1^2 + PC2^2} \end{aligned} \tag{1}$$

In the coordinate system defined by the first three PCs, a spherical surface is assumed with a radius *R* that can be expressed as $R^2 = PC1^2 + PC2^2 + PC3^2$. Based on 24,472 records over the past 46 years, the mean absolute values of PC1, PC2, and PC3 are approximately 130, 42, and 33, respectively. Accordingly, the radius of the sphere *R* can be taken as 140. Following the definitions of ϕ and λ, the corresponding virtual component values are computed as follows:

$$\begin{aligned} PC1 &= R\cos(\phi)\cos(\lambda) \\ PC2 &= R\cos(\phi)\sin(\lambda) \\ PC3 &= R\tan(\phi) \end{aligned} \tag{2}$$

In Eq. (1), ϕ reflects the degree of horizontal asymmetry (EOF3): a positive (negative) ϕ indicates a positive (negative) PC3 value, meaning that the thermal structure exhibits a positively (negatively) phase of horizontal asymmetry mode (EOF3). The larger the absolute value of ϕ, the stronger the horizontal asymmetry in the lower-to-mid levels temperature anomalies. λ represents the relative magnitude of the baroclinic mode (EOF2) to the typical warm core mode (EOF1). When $|\lambda| < 45°$ or $|\lambda| > 135°$, the absolute value of PC1 is greater than that of PC2; otherwise, the absolute value of PC1 is less than that of PC2. In the special case where ϕ =0° and λ=0°, the solution yields PC2=0 and PC3=0, corresponding to the typical warm core structure of EOF1. Conversely, when ϕ = 0° and λ = 90°, the result gives PC1=0 and PC3=0, which represents the baroclinic structure of EOF2, characterized by a warm core in the mid-to-upper levels and cooler air in the lower-to-mid levels.

From the climatological frequency distribution of TCs for TS intensity and above (Fig. 7b), the contour lines show an elliptical shape, with λ showing a wider range of variation than ϕ. Moreover, the heatmap (Fig. 7b) exhibits an asymmetric frequency distribution: high values cluster along the band $-10° \le \phi \le 5°$, $0° \le \lambda \le 20°$, with the maximum located in the single grid box $-5° \le \phi \le 0°$, $10° \le \lambda \le 15°$. To better show the one-dimensional distributions of ϕ and λ, frequency histograms of these parameters were created. The overall shapes of these histograms roughly follow a Gaussian (bell-shaped) curve, but there is asymmetry, with the ϕ curve slightly skewed to the left (Fig. 7c) and the λ curve slightly skewed to the right (Fig. 7d). As shown in Figs. 7c, d, 98.8% of the ϕ values fall within the range of -60° to 60°, while 98.5% of the λ values fall within -60° to 90°. Since the sample data are not widely spread over the whole phase space, in the following discussion, the parameter ranges were set from -60° to 60° for ϕ and -60° to 90° for λ.

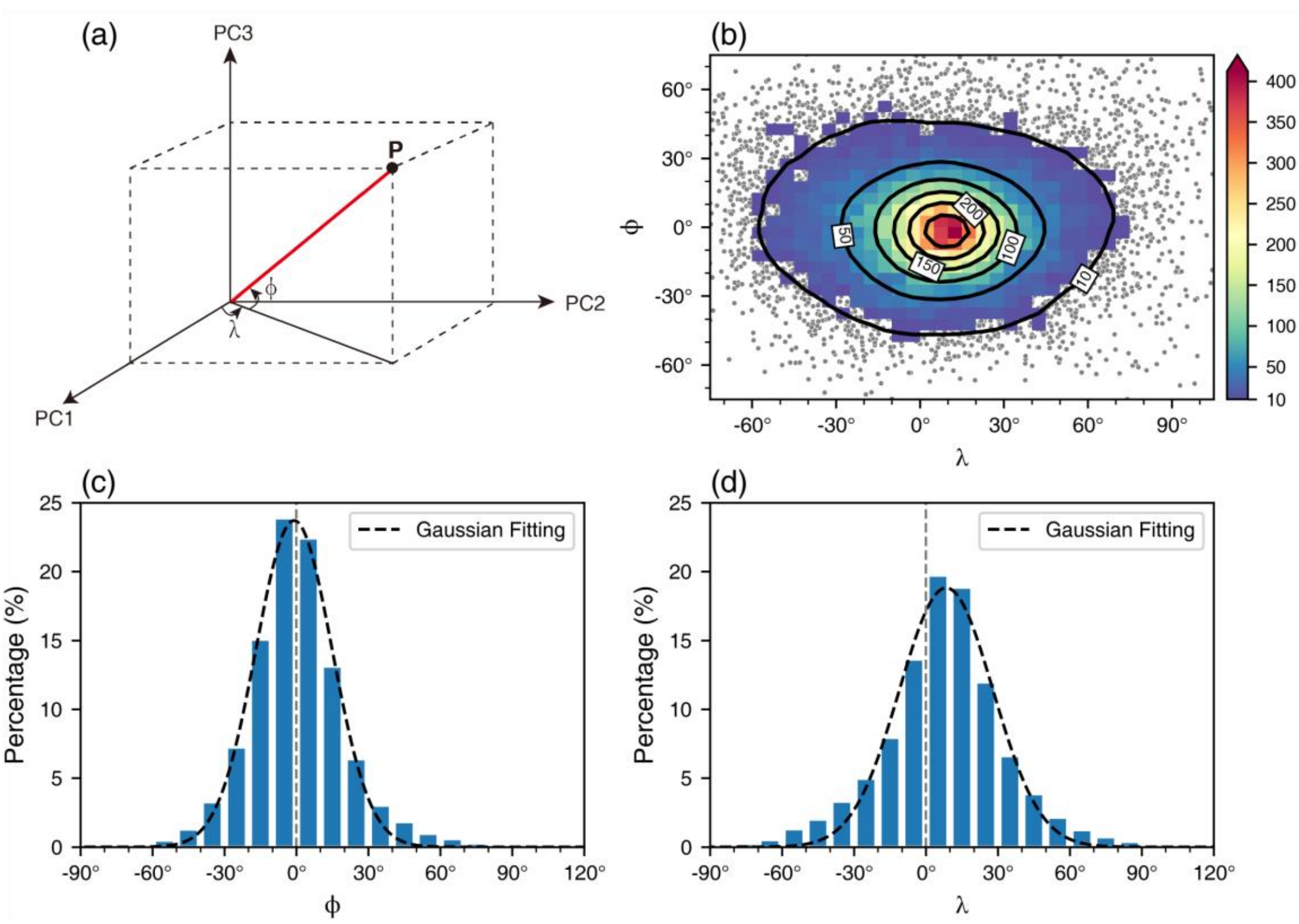

**Fig. 7** (**a**) Schematic diagram of the thermal structure parameters ϕ and λ for TC. (**b**) Scatter plot and frequency heatmap with 5°×5° grid resolution and contours for TS intensity and above during 1979–2024. (**c**) Frequency percentage distribution and Gaussian fitting curve for the parameter ϕ. (**d**) Frequency percentage distribution and Gaussian fitting curve for the parameter λ. The vertical dashed line in (**c**, **d**) indicates the zero value of ϕ and λ, respectively.

### 4.2 Construction of the phase diagram construction

As previously noted, the first three PCs can capture the dominant signals in the temperature anomaly field. Case studies of two TCs further demonstrate that the evolution of the first three PCs closely mirrors the thermodynamic structural evolution of each system throughout its life cycle. Thus, the two phase space parameters (ϕ and λ), derived from the first three PCs, can effectively visualize the status of TC thermal structure and its evolution. In this section, we examine the distribution of $R_{123}$ across all time steps of the temperature anomaly fields to confirm the ability of this pair of parameters to capture a TC's thermal structure. As a rule, the larger $R_{123}$ is, the more dominant the three leading modes are, and the more accurately these two parameters represent the TC's temperature-anomaly structure. As can be seen in Fig. 2, $R_{123}$ spans a wide range among TC categories, with certain values exceeding 90% and some samples dropping below 10%. Across the full sample of 24,472, it is necessary to evaluate the consistency between the temperature anomaly field reconstructed by the first three PCs and the original temperature anomaly field from ERA5. A high overall correlation coefficient indicates that the temperature field reconstructed by the first three PCs not only captures the majority of the variance of the original temperature anomaly but also highly matches the detailed spatial features (such as the location of cold and warm anomaly centers) and temporal evolution, including the magnitude fluctuations of anomalies.

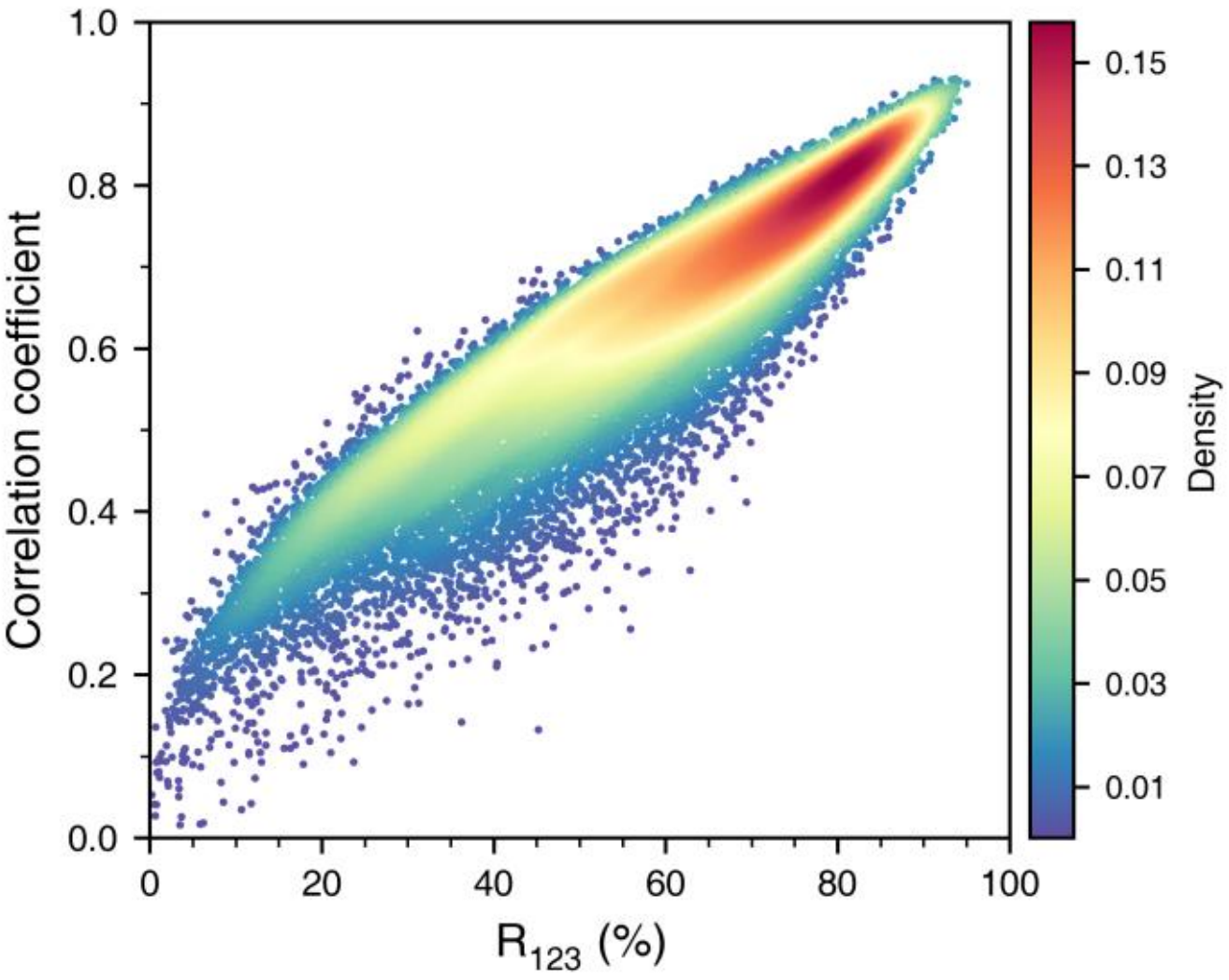

**Fig. 8** The scatter plot of the $R_{123}$ and the correlation coefficient between the temperature anomaly field reconstructed from the first three PCs and the original temperature anomaly field from the ERA5 dataset. The colors represent the density of the scatter points.

Fig. 8 reveals a distinct positive linear relationship between $R_{123}$ and the correlation coefficient comparing the ERA5 original temperature anomaly field with the field reconstructed from the first three PCs. When the $R_{123}$ values are small, the scatter points in the figure are more dispersed. At this time, the temperature anomaly field reconstructed from the first three PCs contains more noise or other high-wavenumber features, and has a larger difference from the original temperature anomaly field from ERA5, resulting in a smaller correlation coefficient between the two fields. In contrast, when the $R_{123}$ values are large, the scatter points are more densely distributed, especially when $R_{123}$ is between 60% and 90%. A larger $R_{123}$ means that the first three PCs can better explain the temperature field variability and have a good ability to reconstruct the temperature anomaly field (high correlation).

In this study, we employ Fisher's z-transformation (Wilks, 2006) to compute mean correlation coefficients (i.e., corresponding z values). The transformation is applied to each correlation coefficient $r$ as

$$z(r) = \frac{1}{2}\ln\left(\frac{1+r}{1-r}\right) \tag{3}$$

We then use the inverse function of Eq. (3) to derive a mean correlation. This is standard practice for averaging correlation coefficients (Pennell and Reichler 2011).

Statistically (Table 1), there are 23,197 samples with $R_{123}$ exceeding the 20% contribution threshold, accounting for 94.8% of the total frequency. The average correlation coefficient between the observed and reconstructed fields of these samples reaches 0.66 ($p<0.01$), which is in an acceptable range in meteorological forecast verification (Dai 2016; Dong et al. 2023; Sun et al. 2024). When the contribution threshold is raised to 50%, there are 15,924 samples (65.1% of the total), and the average correlation coefficient exceeds 0.73($p<0.01$), indicating good reconstruction skill. These results consistently support the conclusion that retaining the first three PCs is an appropriate and effective strategy for reconstructing the dominant thermal structure of tropical cyclones.

**Table 1** Frequency, percentage, mean correlation coefficient, and p-value of $R_{123}$ sub-samples exceeding different $R_{123}$ thresholds. The total sample consists of 24,472 times of TC intensity at TS and above during 1979–2024. The correlation coefficient is defined as in **Fig. 8**.

| | $R_{123}$ threshold | | | | | | |
|---|---|---|---|---|---|---|---|
| | ≥20% | ≥30% | ≥40% | ≥50% | ≥60% | ≥70% | ≥80% |
| Count | 23197 | 21546 | 19066 | 15924 | 11930 | 7391 | 3040 |
| Percentage (%) | 94.8 | 88.0 | 77.9 | 65.1 | 48.7 | 30.2 | 12.4 |
| Mean Corr | 0.66 | 0.68 | 0.70 | 0.73 | 0.75 | 0.79 | 0.84 |
| p-value | <0.01 | <0.01 | <0.01 | <0.01 | <0.01 | <0.01 | <0.01 |

To subjectively evaluate the thermal structure characteristics corresponding to different ϕ and λ values, we here present the temperature anomalies field reconstructed based on ϕ = -60°, -30°, 0°, 30°, 60°, and λ = -60°, -30°, 0°, 30°, 60°, 90°, based on the distributions shown in Fig. 7c-d and Eq. (2).

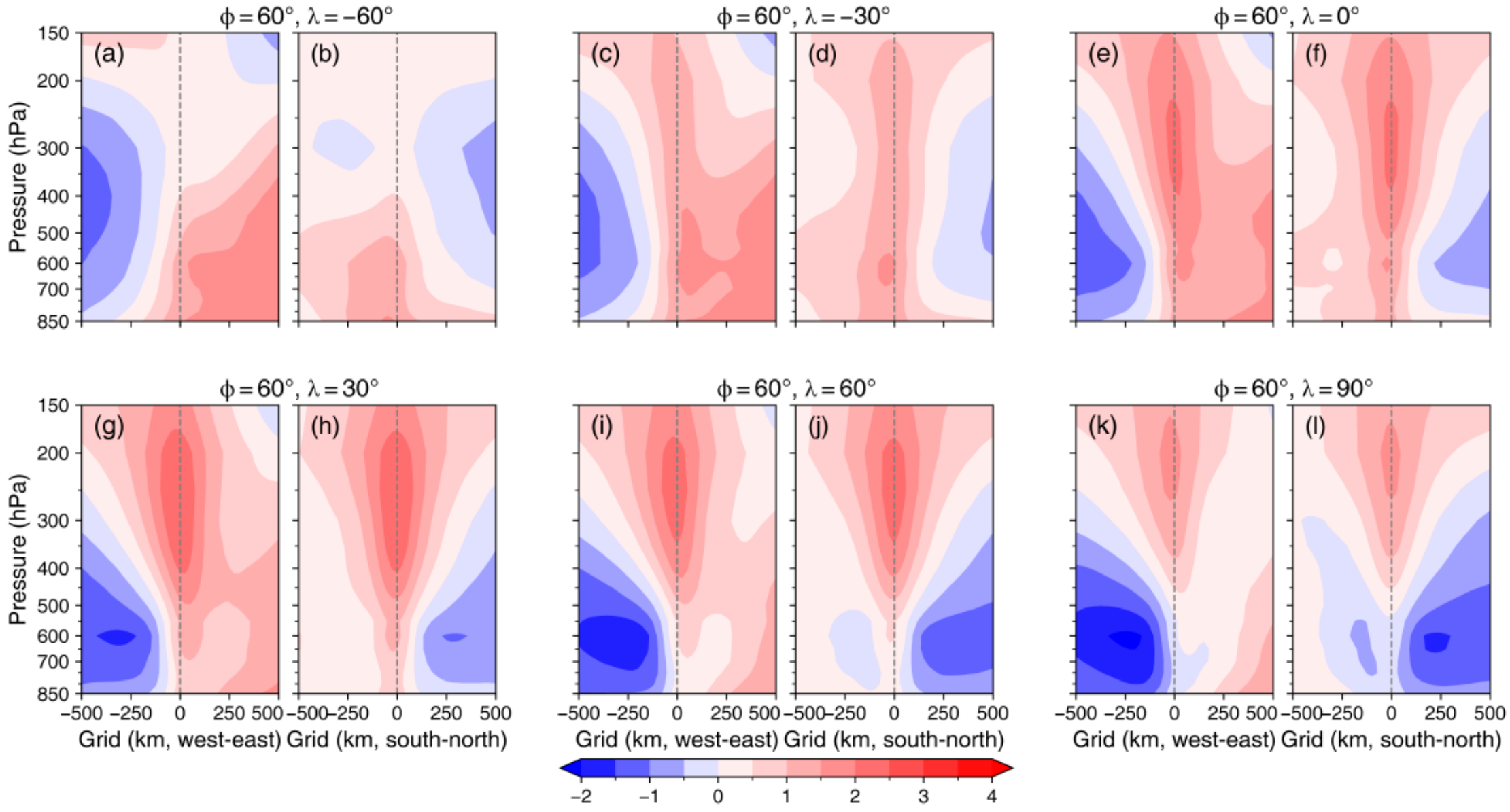

**Fig. 9** Reconstructed temperature anomaly fields (units: °C) for ϕ = 60° and λ values of -60°, -30°, 0°, 30°, 60°, and 90°. Left column (**a, c, e, g, i, k**): zonal-vertical cross-sections; right column (**b, d, f, h, j, l**): meridional-vertical cross-sections. The grey dashed lines represent the TC center from IBTrACS.

Fig. 9 shows the zonal- and meridional-vertical cross-sections of the temperature anomaly fields along the TC center for different λ values when ϕ=60°. A ϕ value of 60° corresponds to a substantial contribution from PC3, reflecting a pronounced horizontal asymmetry in the thermal structure. When λ=-60° (Fig. 9a, b), the positive anomalies are mainly concentrated in the lower-to-mid levels, with negative anomalies appearing aloft in the west and north sides at 300–600 hPa, and the negative anomaly centered around 400 hPa, with the negative anomaly on the west side being stronger than that on the north side. For λ=-30° (Fig. 9c, d), the positive anomaly region extends upward, forming a vertically coherent structure with a weak warm core around 750 hPa, and the negative anomaly area descends vertically to 400–700 hPa, with the negative anomaly center around 500 hPa. Under λ=0° (Fig. 9e, f), the positive anomaly peaks in the mid-to-upper levels, with the warm core located at 250–400 hPa, consistent with the traditional height of the TC warm core. Meanwhile, the positive anomaly area on the east side of the TC still persists in the lower-to-mid levels, and the negative anomalies expand in the lower-to-mid levels, gradually intruding into the TC center. For λ=30° (Fig. 9g, h), the positive anomalies in mid-to-upper levels become more concentrated, but negative anomaly centers can be seen on the west and north sides at 600 hPa, with stronger negative anomalies in the west. For λ=60° (Fig. 9i–j), the negative anomaly in mid-levels intensifies and spreads horizontally, with negative anomalies extending into the southern flank, although the TC center at 850 hPa remains a positive anomaly. For a higher λ of 90° (Fig. 9k, l), negative anomalies expand and dominate throughout the lower-to-mid levels, completely enveloping the TC center at 500–850 hPa, TC center is entirely occupied by the negative anomalies. In summary, for ϕ = 60°, as λ increases from -60° to 90°, the thermal structure evolves systematically: negative anomalies in mid-levels originating from the northwest sector progressively deepen and intensify, eventually eroding the warm core, while the positive anomalies center shift from the lower-to-mid levels to the mid-to-upper levels before being overtaken by negative anomalies.

Reducing ϕ from 60° to 30° diminishes the role of PC3 and increases the influence of PC1 and PC2 in simulating the temperature anomaly field. Fig. 10 shows that the evolution of the temperature anomaly fields at ϕ=30° are similar to that in Fig. 9, but with a stronger warm core and more intense negative anomalies in the lower-to-mid levels.

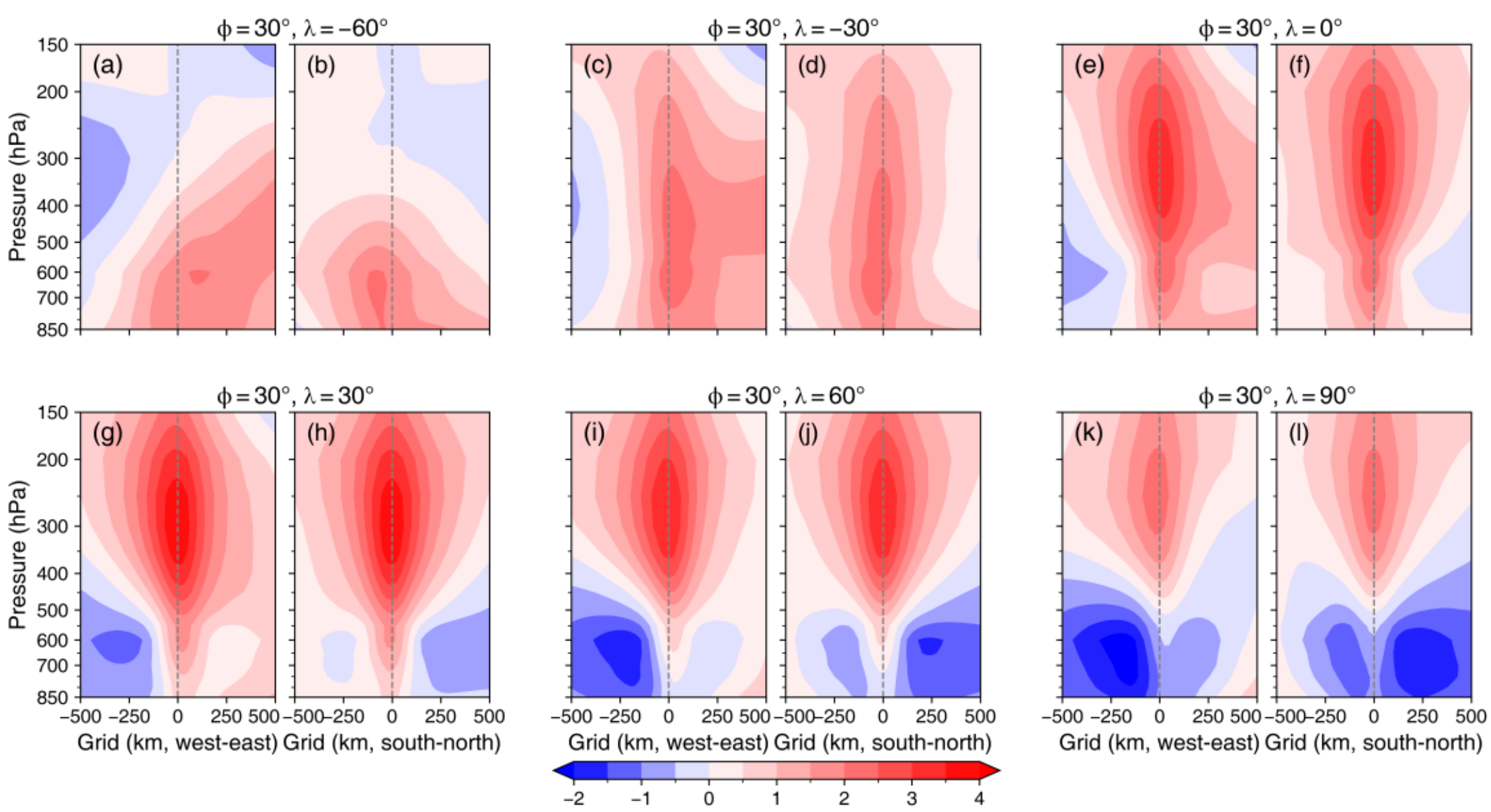

**Fig. 10** The same as **Fig. 9**, but for ϕ = 30°.

Notably, PC3 becomes zero when ϕ=0°, and the temperature anomaly field is entirely simulated by the two modes of EOF1 and EOF2. Compared to other values of ϕ, the simulated temperature anomaly field at this time will exhibit more symmetric horizontal structure (Fig. 11). Specifically, at λ = -60°, which shows distinct negative anomaly in mid-to-upper levels and a warm core at 600–850 hPa, all other λ values exhibit positive anomalies aloft with the warm core shifting upward (Figs. 11e, l). At λ=30°, where both PC1 and PC2 are positive, the influence of EOF2 produces notable negative anomalies in the lower-to-mid levels about 100 km from the TC center (Figs. 11g, h). As λ increases further, the negative anomaly is observed to strengthen while remaining consistently located approximately 250 km from the TC center. In the specific case of λ = 0° (90°), PC2 (PC1) becomes zero, respectively. Consequently, the temperature anomaly field becomes proportionally similar to the spatial pattern of EOF1 or EOF2 alone.

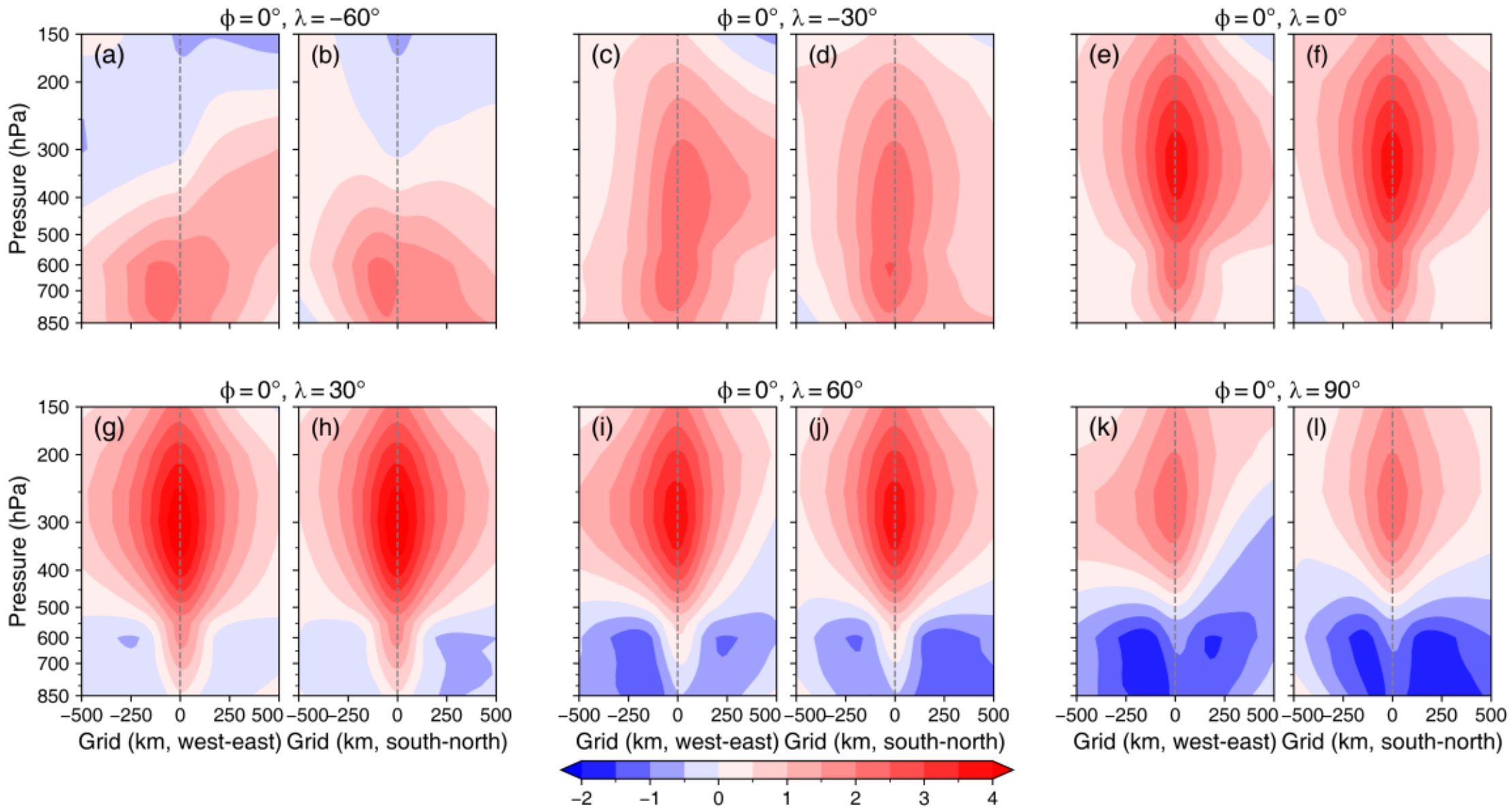

**Fig. 11** The same as **Fig. 9**, but for ϕ = 0°.

With ϕ fixed at -30° (indicating a negative PC3), the reconstructed temperature anomaly field (Fig. 12) shows that, in contrast to the ϕ=30° case, negative temperature anomaly in lower-to-mid levels initially appears in the southeastern sector. As λ increases, this negative anomaly gradually intrudes into the northwestern part of the TC, while the warm core shifts upward from the lower-to-mid levels to the mid-to-upper levels. A similar evolution is observed for ϕ=60° (Fig. 13), where the horizontal asymmetry of the temperature anomalies is further enhanced, and the warm core maximum becomes displaced from the TC center.

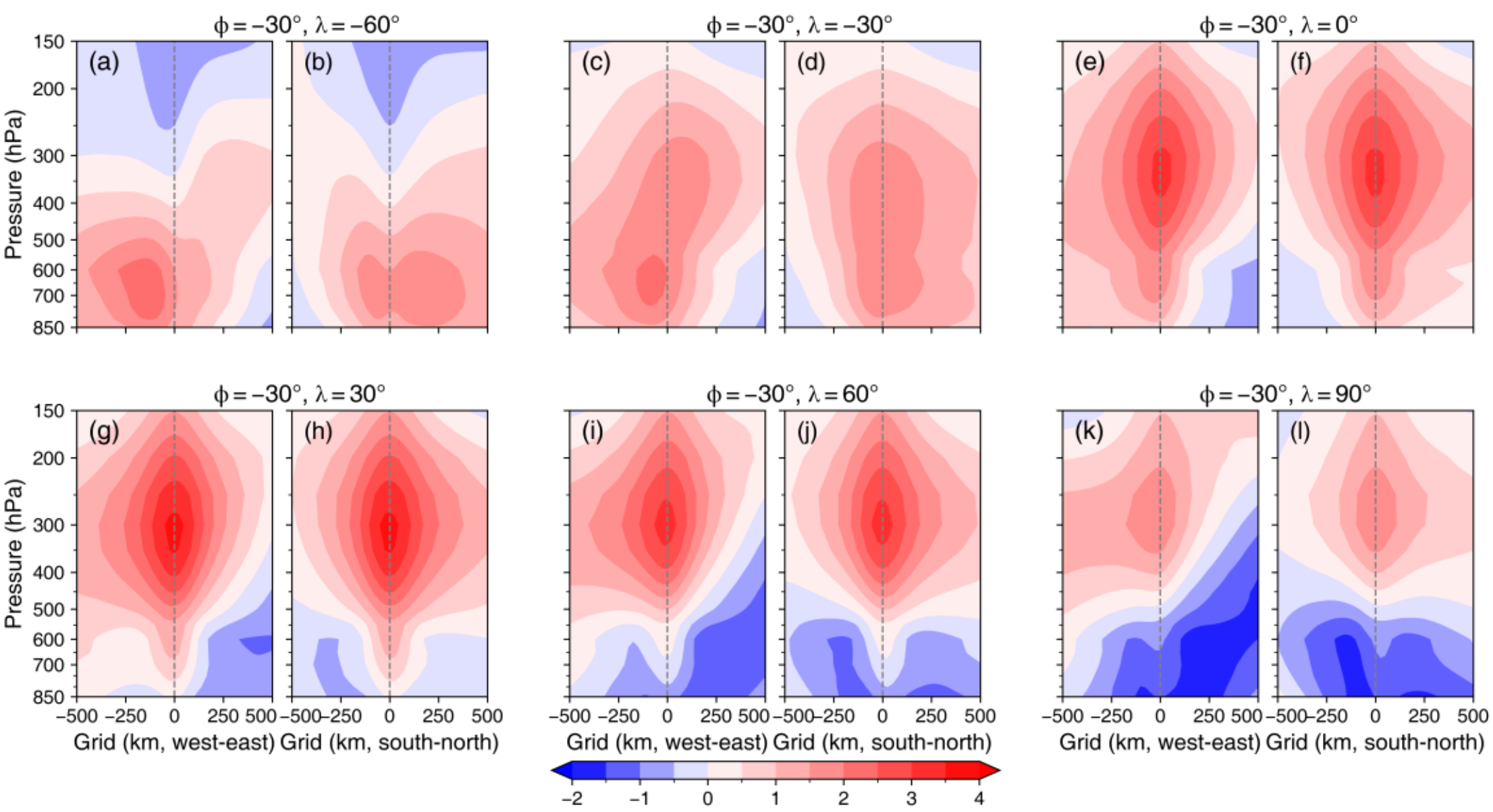

**Fig. 12** The same as **Fig. 9**, but for ϕ = -30°.

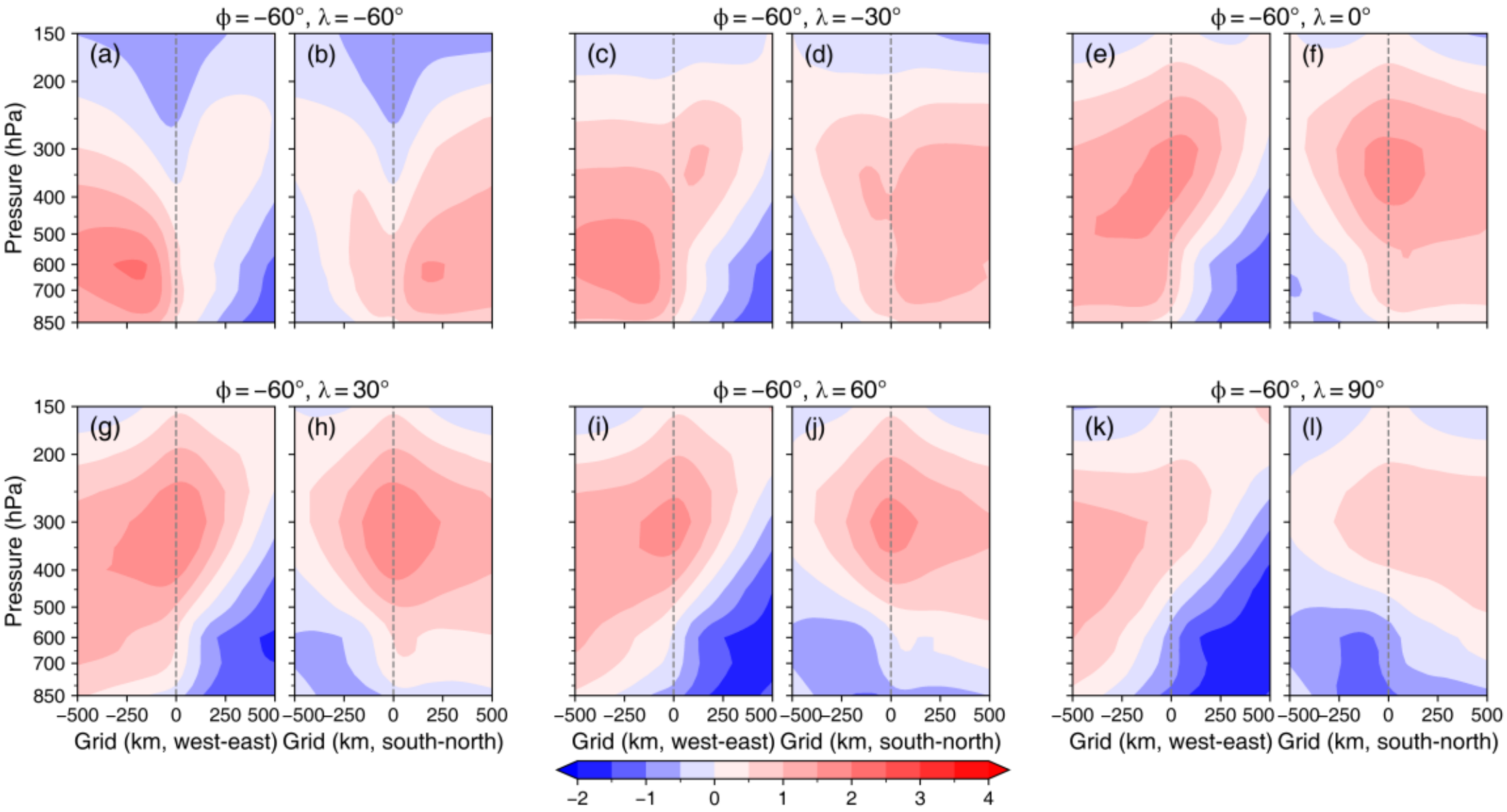

**Fig. 13** The same as **Fig. 9**, but for ϕ = -60°.

From the figures above, it can be seen that different ϕ and λ values can effectively capture key features of the thermal structure, including the spatial extent of positive anomalies, the altitude, and intensity of the warm core. Furthermore, the height and intensity of the warm core, as well as the initial height, expansion direction, and horizontal asymmetry of the negative anomalies, can also be observed. Thus, one can construct a ϕ–λ phase diagram, in which each point in the diagram corresponds to the thermal structure of the TC and the structural changes in its life cycle.

To make it clearer, Fig. 14 provides a schematic diagram of the thermal phase space in which the vertical structures can be subjectively grouped into five categories (Fig. 14a): (1) warm core in the lower-to-mid levels, (2) warm core in the mid levels, (3) warm core in the mid-to-upper levels, (4) coexistence of warm core in mid-to-upper levels and cold core in the lower-to-mid levels, and

(5) coexistence of cold core in the mid-to-upper levels and warm core in the lower-to-mid levels. Regarding the horizontal asymmetry of positive and negative anomalies, when ϕ>0°, the positive (negative) anomaly is located on the southeast (northwest) sector. As λ increases from -60° to 90°, the negative anomaly in the mid-to-upper levels gradually shifts to the lower-to-mid levels. When ϕ < 0°, the negative anomaly in the mid-to-upper levels weakens, while the negative anomaly in the lower-to-mid levels on the southeast sector increases and intrudes into the TC center. Based on this asymmetry of positive and negative anomalies, the patterns can be subjectively classified into four categories: negative anomalies in the mid-to-upper levels, mid levels, lower-to-mid levels, and coexistence of negative anomalies in the mid-to-upper levels and lower-to-mid levels (Fig. 14b).

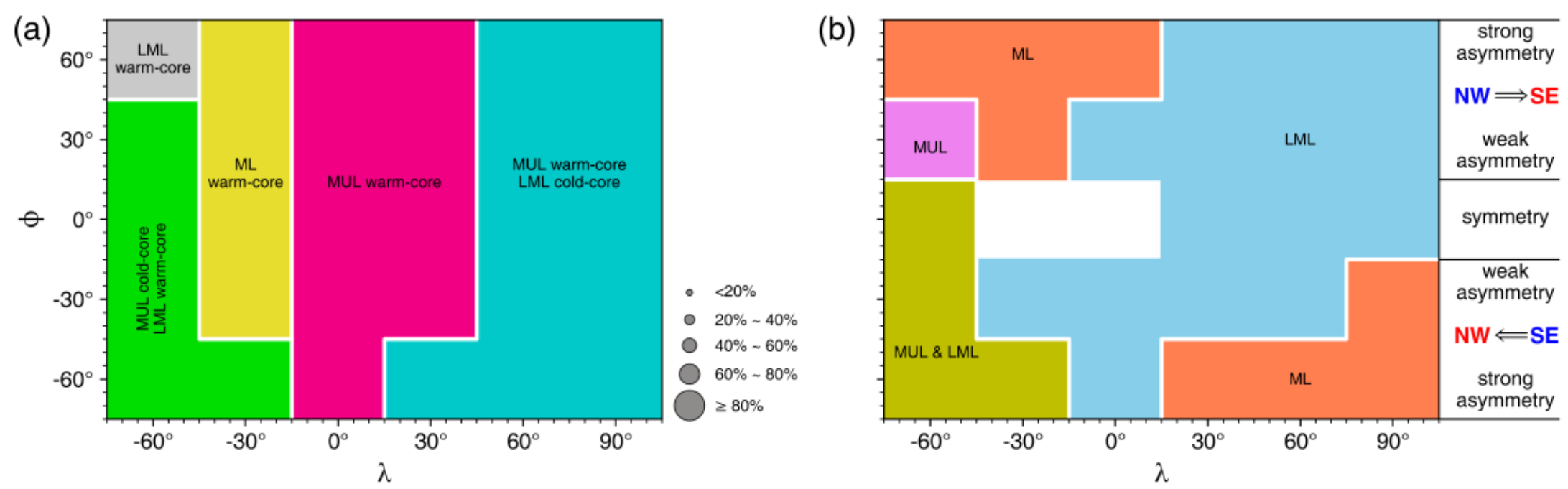

**Fig. 14** Schematic diagram of TC thermal structures in the ϕ–λ phase space: (**a**) vertical distribution of temperature anomalies, and (**b**) spatial distribution of negative anomalies and associated horizontal asymmetry. Abbreviations in denote vertical layers: LML (lower-to-mid levels, 600–850 hPa), ML (mid levels, 400–600 hPa), and MUL (mid-to-upper levels, 150–400 hPa). The size of the gray dots in (**a**) is proportional to the value of $R_{123}$. In the diagram, labels "NW" and "SE" in (**b**) denote northwest and southeast regions relative to the TC center, while red and blue fonts indicate positive and negative anomalies, respectively. The blank areas indicate the absence of negative anomalies.

### 4.3 Applications of the TC thermal phase space

In this subsection, we demonstrate how the TC thermal phase space introduced above can be applied to TC analyses. Based on the two TCs mentioned above, we plot their life cycle trajectories onto the thermal phase diagram to track how their thermal structures evolve with intensity and to demonstrate the diagram's capacity for intuitive diagnosis of TC thermal anomalies.

Fig. 15a shows the evolution of the vertical thermal anomaly structure of Chaba (2022). At 1200 UTC on 29 June 2022, when Chaba was at TD intensity, ϕ = -5.3° and λ = 11.1°, which corresponds to a weak thermal asymmetry, is consistent with Chaba's relatively weak and less organized warm anomaly concentrated in the mid-to-upper levels (Fig. 4a–d). Meanwhile, the value of $R_{123}$ was 65%, exceeding the average value of $R_{123}$ for the TD stage (Fig. 2). The correlation between the anomaly field reconstructed from the first three PCs and the original anomaly field in ERA5 was between 0.65 and 0.70 (Fig. 8, Table 1), indicating that the three leading PCs well captured the dominant features of the original temperature anomaly field. When the TC reached TS intensity, more vigorous deep convection led to a more concentrated mid-to-upper levels warm core. This structural change resulted in a greater dominance of PC1 and a consequent decrease in λ to 5°. At this time, $R_{123}$ exceeded 80%, indicating that the three dominant modes provide an excellent representation of the temperature anomaly structure. The observed increases in the magnitudes of both ϕ and λ from July 1 to 2 reflect the warm core development (increasing PC1) and increased lower-to-mid levels thermal asymmetry (enhancing negative PC3), a relationship clearly shown in Figs. 3c-e. At 0600 UTC on 3 July, Chaba was at its last time of TS intensity, it exhibited a warm core in mid-to-upper levels and negative anomaly in lower-to-mid levels, shortly weakened to TD intensity. At this time, the larger value of PC2 compared to PC1 indicated that negative anomalies prevailed in the lower-to-mid levels, which was associated with a λ value exceeding 60°.

In terms of horizontal asymmetry of anomalies (Fig. 15b), the initial stage was dominated by PC1 with no notable negative anomalies near the center, corresponding to very small values of ϕ and λ in the phase space. When λ began to increase at 1800 UTC on 1 July, the point in phase diagram entered the LML region, indicating the emergence of negative anomalies in lower-to-mid levels, which is clearly shown in Fig. 3e. Throughout the period when negative anomalies in lower-to-mid levels were present, ϕ was always negative, which also means that the negative anomaly first appeared in the southeast sector. As shown in Fig. 15, the TC phase trajectory closely mirrors the evolution of Chaba's thermal structure.

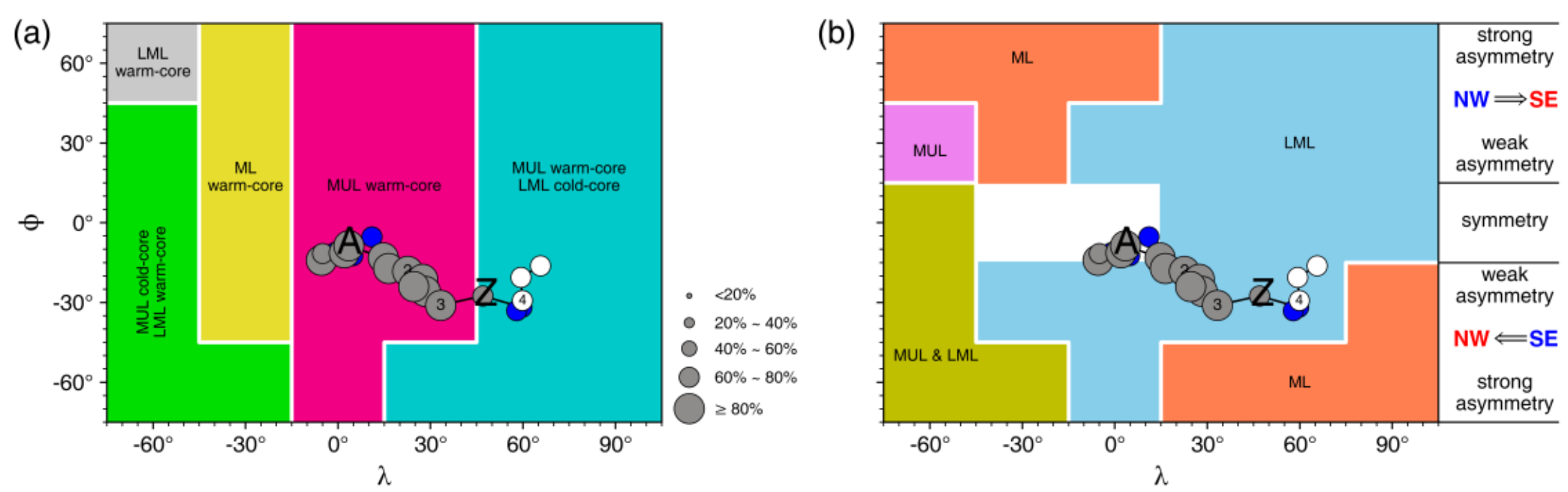

**Fig. 15** Example application of thermal phase diagram for Typhoon Chaba (2022). The A and Z denote the first and last times, respectively, when the storm intensity reached or exceeded TS (black dashed lines in Fig. 3b). Point colors indicate intensity stages: gray for TS or stronger, blue for TD, and white for extratropical cyclone (EX) or low(LO). Other symbols follow the same conventions as in **Fig. 14**.

For Kong-rey (2024), it exhibited a different trajectory in the thermal phase space compared to Chaba. Since the warm core intensity during the initial TD stage of Kong-rey's life cycle was relatively weak (Fig. 6a), the $R_{123}$ value was only 48% (Fig. 16a). The vertical height of the warm core varied considerably because there was a lack of sustained convection near the cyclone center when TC is weak. During this stage, λ also varied sharply, which may be correlated with the fact that the temperature anomaly field was linked to higher-order principal components, not the first three. After reaching TS intensity at 1200 UTC on 25 October, PC1 gradually increased and became dominant, and the warm core stabilized in the mid-to-upper levels, so the values of ϕ and λ are close to 0°, accompanied by an $R_{123}$ value exceeding 80%. At 0000 UTC On 30 October, Kong-rey reached its LMI (Category 5), as the $R_{123}$ also peaked at 85%. On 1 November, influenced by an eastward-moving mid-latitude trough to the northwest, negative temperature anomalies emerged in mid-to-upper levels and PC2 turned negative (Fig. 5d), while the warm core descended to the mid levels. The increased horizontal asymmetry in the thermal structure corresponds to a rising ϕ value, and the negative temperature anomaly in the mid-to-upper levels caused λ to shift from positive to negative values. Following extratropical transition, Kong-rey finally transformed into an extratropical cyclone, and a cold core developed in the mid-to-upper levels, though a warm core persisted in the lower-to-mid levels. The disorganized thermal anomalies in the low-to-mid levels resulted in a small ϕ value, while the pronounced cold anomaly in the mid-to-upper levels maintained a large magnitude for λ. This vertical evolution of thermal structures well matched with TC formation, intensification, and interaction with the mid-latitude trough, and phase space can well visualize the evolution of Kong-rey's thermal structure.

In the changes of negative temperature anomalies, since Kong-rey was generated over the vast tropical ocean, the negative anomalies in the lower-to-mid levels were very small before LMI, and the temperature anomaly field was nearly axisymmetric (with very small ϕ). On 30 June, ϕ remained positive and began to increase, indicating that the asymmetry of the temperature anomaly field increased during the decay stage, and negative anomalies began to appear in the northwest. On 1 July, the negative anomalies in the mid levels in the northwest became more significant, and one day later, the mid-to-upper levels were fully occupied by cold air after Kong-rey transformed into an extratropical cyclone.

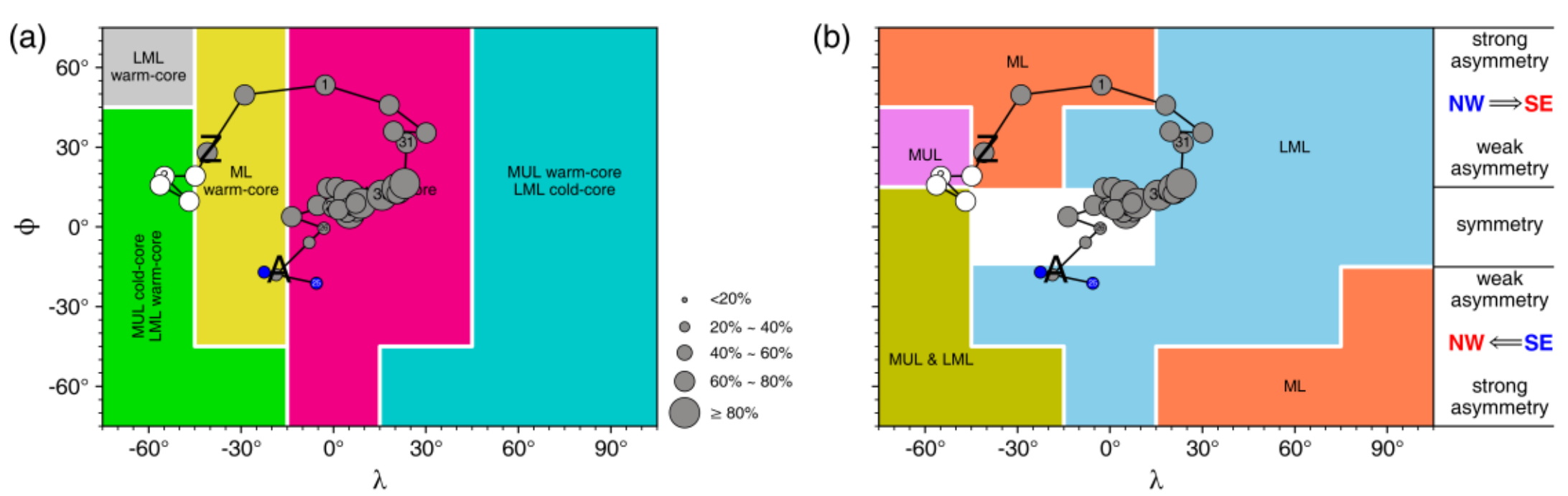

**Fig. 16** The same as **Fig.15**, but for Kong-rey (2024).

To further explore the relationship between TC intensity changes and thermal phase, we summarized their correlation from 1979 to 2024 in Table 2. In terms of climatology, the highest proportion of warm core appears in the mid-to-upper levels, with an average of 81.98%. This is followed by a warm core confined to the mid levels (9.75%). The coexistence of warm core in mid-to-upper levels and cold core in the lower-to-mid levels accounts for 6.58%, while the opposite configuration, the coexistence of cold core in mid-to-upper levels and warm core in lower-to-mid levels accounts for 1.54%. The proportion of warm core within the mid levels is the smallest, at only 0.15%.

For the different TC stages, a distinct pattern emerges: only when the warm core appears in the mid-to-upper levels do the two types of intensification (slowly intensifying, SI and rapid intensifying, RI) have a higher proportion than neutral (N) and weakening (W) types (Table 2), this phenomenon is consistent with the conclusions drawn from numerical simulations and satellite data analyses of tropical cyclones undergoing RI (Kieu et al. 2016; Shi et al. 2021; Gao et al. 2017; Wang and Jiang 2019). A higher warm core altitude is hydrostatically more efficient for surface pressure decrease, indicating that eye warming in the upper troposphere is more critical to rapid intensification. (Chen and Zhang 2013; Wang and Wang 2014; Miller et al. 2015). In all other cases, the proportion of neutral (N) and weakening (W) types is higher than the other two types. Notably, a warm core located in the lower-to-mid levels never coincides with RI.

**Table 2** Distribution of thermal phase space types under different intensity change categories。Intensity changes

categories: weakening (W), neutral (N), slowly intensifying (SI), and rapid intensifying (RI).

| | Category (%) | | | | Climatology (%) |
|---|---|---|---|---|---|
| | W | N | SI | RI | |
| LML warm-core | 0.31 | 0.25 | 0.03 | 0.00 | 0.15 |
| ML warm-core | 12.10 | 13.02 | 8.64 | 5.25 | 9.75 |
| MUL warm-core | 78.15 | 76.92 | 84.01 | 88.84 | 81.98 |
| MUL warm-core & LML cold-core | 6.83 | 7.45 | 6.55 | 5.49 | 6.58 |
| MUL cold-core & LML warm-core | 2.61 | 2.36 | 0.77 | 0.43 | 1.54 |

## 5 Discussion and conclusions

This study attempts to explore the dominant TC thermal structure for TS intensity and above over the WNP basin. Using the first three principal components (PCs) corresponding to the EOF modes, we established a three-dimensional Cartesian coordinate system. Within this framework, we introduce the TC thermal phase space to quantify the TC's three-dimensional thermal structure: one quantifies the horizontal asymmetry of the thermal field, and the other diagnoses the vertical barotropic or baroclinic structure. Two recent cases were used to illustrate that the three PCs time series, as well as the TC thermal phase space, correspond well with observed intensity changes and thermal structural evolution. Therefore, the EOF-based thermal phase space provides a novel and objective tool for analyzing and diagnosing structural dynamics characteristics and intensity evolution.

Rather than using the geopotential height fields employed in the CPS approach, we utilize a temperature dataset from the ERA5 dataset to depict the evolution of TC the thermal structure. In the CPS framework, the warm and cold cores are inferred from geopotential height via the thermal wind balance. However, several studies have shown that the wind fields in the ERA5 dataset are often over-smoothed when depicting TC structure, leading to an underestimation of maximum wind speeds (Xiong et al. 2022; Liu et al. 2025; Ye et al. 2026). This inaccuracy consequently propagates to the balanced height field (minimum central pressure). In contrast, the temperature field is less affected by sub-grid-scale processes and exhibits more coherent spatial variability in models (Murillo et al. 2022). Furthermore, the temperature field is strongly constrained by the assimilation of satellite and radiosonde data (Song et al. 2024). For these reasons, the temperature field generally offers a more coherent and reliable representation of the TC thermal structure, particularly for characterizing the warm core.

As for the temperature anomaly field, it was derived through a two-step procedure to ensure a clean separation of the TC's intrinsic thermal signal. First, for each observation time of the TC, the original ERA5 temperature field was subtracted by the long-term mean (1991–2020) for the same calendar day and month from the original ERA5 temperature. This step effectively removes the slowly varying, latitude-dependent signals produced by seasonal solar radiation and sea surface temperature, ensuring that climatic gradients are not misinterpreted as part of TC-related signals. Second, the annular mean computed within a 600−1000 km radius of the TC center was also subtracted. This ring lies just outside the outermost closed circulation, so it contains the immediate thermodynamic environment (such as baroclinic zones, high-level troughs) while excluding the eyewall, inner rainbands. After these two adjustments, the temperature anomaly field exhibits smooth temporal evolution throughout the TC’s life cycle, without abrupt changes, and is well suited to EOF analysis or filtering techniques (Hannachi et al. 2007). Ultimately, this yield results that are both more reliable and more readily comparable across cases.

The EOF results show that the first leading EOF mode is characterized by a vertically coherent, consistent positive anomaly within 500 km of the TC center, with the warm core located between 250 and 400 hPa, which is a typical TC warm core structure. The second EOF mode is marked by baroclinicity vertically, with opposite temperature anomalies above and below 500 hPa. The upper warm core sits directly above the TC center at 250–300 hPa, whereas the lower cold core is displaced ~100 km outward from the center. The second EOF pattern captures the intrusion of mid-latitude upper-level trough cold air into TC or the ingress of relatively colder air into the TC circulation in the lower levels. The third EOF mode represents the horizontal asymmetry of temperature anomalies, which can reflect the intensity and the expanding direction of negative temperature anomalies affecting the TC.

We examined the PCs time series for TCs Chaba (2022) and Kong-rey (2024), the variations in PCs closely correspond to intensity changes and the evolution of the temperature anomaly field. Since the EOF decomposition was performed based on the temperature anomaly fields of relatively strong TCs (TS and above), PCs obtained by projecting the temperature anomaly fields at TD and EX stages onto the EOF modes may not always well reflect the true thermal state at these stages. However, they can still serve as meaningful indicators of whether the thermal structures during the TD and EX stage already resemble the temperature anomaly patterns characteristic of strong TCs, as quantified by $R_{123}$.

In summary, EOF analysis effectively extracts the most dominant modes of TC thermal structures and their corresponding PCs from vast amounts of TC temperature data. By constructing a phase space diagram with the first three PCs, the complex, high-dimensional, and continuous evolution of the TC's thermal structure is distilled into low-dimensional parameters that have clear physical meaning and form a coherent trajectory in the phase space. Using the two angles $\phi$ and $\lambda$ derived from this trajectory, multiple structural characteristics, such as symmetric vs. asymmetric patterns, deep vs. shallow warm cores, and the vertical height of the warm core, can be quantified. Unlike the CPS method, this approach does not require determining the TC's moving direction; consequently, the PC curves are comparatively smoother and more continuous, effectively capturing the coherence of the TC's thermal evolution and making structural differences across various lifecycle stages both intuitive and rigorous. Future research could explore the relationship between $\phi$ (horizontal asymmetry) and vertical wind shear, and investigate the characteristic range of $\lambda$ (warm core intensity) during RI. Moreover, the parameters $\phi$ and $\lambda$ serve as good indicators that can be integrated into operational TC structure monitoring and forecasting systems. This phase space framework serves as a powerful diagnostic tool, offering profound insight into the dominant physical processes throughout a TC's life cycle and demonstrating considerable promise for both research and operational applications.

**Acknowledgements** This work was supported by the National Natural Science Foundation of China (42075053, 42405038). We thank for the technical support of the National Large Scientific and Technological Infrastructure "Earth System Numerical Simulation Facility" (https://cstr.cn/31134.02.EL).

**Author contributions** Yaoming Ma, Banglin Zhang, and Jeremy Cheuk-Hin Leung conceived the study. The methodology was designed by Jimin Liu, Banglin Zhang, and Jeremy Cheuk-Hin Leung. Jimin Liu and Hong Huang developed analysis codes. The first draft of the manuscript was written by Jimin Liu, and all authors reviewed and edited the manuscript. All authors read and approved the final manuscript.

**Funding** National Natural Science Foundation of China (42075053, 42405038).

**Data availability** All data used in this study are freely available online. The WNP TC best track data provided in IBTrACS are downloaded from https://www.ncei.noaa.gov/data/international-best-track-archive-for-climate-stewardship-ibtracs/v04r01/access/. The fifth generation European Centre for Medium-Range Weather Forecasts (ECMWF) global atmospheric reanalysis (ERA5) is available at https://cds.climate.copernicus.eu/datasets/reanalysis-era5-pressure-levels?tab=download.

## Declarations

**Conflict of interest** The authors declare no conflict of interest.